\documentclass[a4paper,fleqn,usenatbib]{mnras}

\usepackage{newtxtext,newtxmath}
\usepackage[T1]{fontenc}
\usepackage{ae,aecompl}

\usepackage{amsmath,amssymb}
\usepackage{graphicx}
\usepackage{latexsym,color}
\usepackage{mathtools,cuted}
\usepackage{natbib}
\usepackage{graphicx}	% Including figure files
\usepackage{amsmath}	% Advanced maths commands
\usepackage{amssymb}	% Extra maths symbols

\title[Multi magnetized accreting tori]{Influence of toroidal magnetic field in multi-accreting  tori}

\author[D. Pugliese et al.]{
D. Pugliese,$^{1}$\thanks{E-mail: d.pugliese.physics@gmail.com}
G. Montani,$^{2}{}^{3}$
\\
$^{1}$Institute of Physics, and Research Centre of Theoretical Physics and Astrophysics, \\
Faculty of Philosophy SIGMA Science,Silesian University in Opava,\\
 Bezru\v{c}ovo n\'{a}m\v{e}st\'{i} 13, CZ-74601 Opava, Czech Republic\\
$^{2}$ENEA- R.C. Frascati, UTFUS-MAG, Via Enrico Fermi 45, Frascati, Roma 00044, Italy\\
$^{3}$Physics Department, ``Sapienza'' University of Rome, P.le Aldo Moro 5, Roma 00185, Italy
}

\date{Accepted XXX. Received YYY; in original form ZZZ}

\pubyear{2015}

\begin{document}
\label{firstpage}
\pagerange{\pageref{firstpage}--\pageref{lastpage}}
\maketitle

\begin{abstract}
We analyzed the  effects of a toroidal magnetic field in the formation of  several magnetized   accretion tori, dubbed  as ringed accretion disks (\textbf{RADs}), orbiting around one central Kerr supermassive Black Hole (\textbf{SMBH}) in \textbf{AGNs}, where
both corotating and counterotating disks are considered.
Constraints on tori formation and emergence of \textbf{RADs} instabilities,   accretion  onto  the central attractor and   tori collision emergence, are investigated.
The results of this analysis show that the role of the central \textbf{BH} spin-mass ratio, the magnetic field and the relative fluid rotation  and tori rotation  with respect the central \textbf{BH}, are crucial elements in determining the accretion tori features,   providing ultimately  evidence of a strict correlation between \textbf{SMBH} spin,  fluid rotation  and magnetic fields in \textbf{RADs} formation and evolution. More specifically we proved that  magnetic field and disks rotation are in fact strongly  constrained, as  tori  formation and evolution in \textbf{RADs} depend on the  toroidal magnetic fields parameters. Eventually  this analysis identifies   specific classes of  tori, for  restrict  ranges of magnetic field parameter, that can be observed    around  some specific \textbf{SMBHs} identified by their dimensionless spin.
\end{abstract}


%%%%%%%%%%%%%%%%%%%%%%%%%%%%%%%%%%%%%%%%%%%%%%%%%%

%%%%%%%%%%%%%%%%% BODY OF PAPER %%%%%%%%%%%%%%%%%%

\def\be{\begin{equation}}
\def\ee{\end{equation}}
\def\bea{\begin{eqnarray}}
\newcommand{\cc}{\mathrm{C}}
\newcommand{\jj}{\mathrm{J}}
\def\eea{\end{eqnarray}}
\newcommand{\tb}[1]{\textbf{\texttt{#1}}}
\newcommand{\ttb}[1]{\textbf{#1}}
\newcommand{\rtb}[1]{\textcolor[rgb]{1.00,0.00,0.00}{\tb{#1}}}
\newcommand{\btb}[1]{\textcolor[rgb]{0.00,0.00,1.00}{\tb{#1}}}
\newcommand{\pp}{\textbf{()}}
\newcommand{\non}[1]{{\LARGE{\not}}{#1}}
\newcommand{\mso}{\mathrm{mso}}
\newcommand{\Qa}{\mathcal{Q}}
\newcommand{\mbo}{\mathrm{mbo}}
\newcommand{\il}{~}
\newcommand{\rc}{\rho_{\ti{C}}}
\newcommand{\dd}{\mathcal{D}}
\newcommand{\Sie}{\mathcal{S}}
\newcommand{\Sa}{\mathcal{S}}

\newcommand{\Mie}{\mathcal{M}}

    \newcommand{\oo}{\mathrm{O}}

\section{Introduction}\label{Sec:wolv}
Magnetic fields are ubiquitous in the Universe, playing a relevant role  in the High Energy  Astrophysics,
and  being involved in a broad variety of precesses in several   environments,  from the early Universe, to solar corona and interstellar medium,
  or in the   Galaxy and galaxy clusters-formation processes,  in Pulsars and  Magnestars.
  In many of these situations however the exact role and  the origin of the  magnetic fields are still to the sorted in a comprehensive picture--see for example
\citep{ER,Siegel:2013nrw,Colgate:2000gb,Grasso:2000wj,Balbus,BHawley}.
The  magnetic field presence  in the  galactic Black Hole (\textbf{BH}) accretion disk environments is a special and intriguing topic.
 The scenario envisaged by the special situation of \textbf{BH} accretion, disk formation, with a  conjectured  accretion-jet correlation is  extremely complex. These issues are  in fact still very much  debated as often correlated with several problematic inherent the  most profound aspects of the \textbf{BH} physics.
In this article we explore the toroidal magnetic fields influence in the accretion  tori  formation,  their configurations  especially in the  emergence of the accretion phase.
More specifically, the analysis focuses     on the   magnetized tori orbiting   super-massive Kerr Black Hole (\textbf{SMBH}) in galactic nuclei (\textbf{AGNs}).
An accretion disk is essentially regulated by the balance of different factors as the  gravitational, centrifugal and magnetic components. In this work we consider clusters of  toroidal  (thick disk) configurations  centered on a single Kerr \textbf{BH}, and prescribed by  barotropic models, for which the time-scale of the dynamical processes $\tau_{dyn}$ (regulated by the gravitational and inertial forces) is much lower than the time-scale of the thermal ones $\tau_{therm}$   (heating, cooling processes and  radiation), that is lower than the time-scale of the viscous processes $\tau_{\nu}$, or $\tau_{dyn}\ll\tau_{therm}\ll\tau_{\nu}$. Consequently the effects of strong gravitational fields are generally dominant with respect to the  dissipative ones and predominant to determine  the systems unstable phases \citep{abrafra,pugtot}.
Each torus is then part of the  coplanar axis-symmetrical structured toroidal disks, orbiting in the equatorial plane of a single central Kerr \textbf{BH},
so called ringed accretion disks (\textbf{RADs}),  introduced in\citet{pugtot} and  detailed in \citep{ringed,open,dsystem,letter,long}.
The \textbf{RADs} model follows the possibility that more  accretion orbiting configurations  can form  around very compact objects  in the special environment of the \textbf{AGNs}-\textbf{BHs} and Quasars. Arising from different \textbf{BHs} accretion periods  and from the   host Galaxy life,  such configurations can report, in their characteristics, traces  of the different periods   during several accretion regimes occurred in the lifetime of non-isolated  Kerr \textbf{BHs} {\citep{Aligetal(2013),Blanchard:2017zfe,long,NixonKing(2012b)}.
During  the  evolution of   \textbf{BHs} in these environments both  corotating  and counterrotating accretion stages are
mixed   during various accretion periods  of the attractor  life \citep{Lovelace:1996kx,Carmona-Loaiza:2015fqa,Dyda:2014pia,Volonteri:2002vz},  thus \textbf{RADs} tori  may be even  misaligned \citep{Aly:2015vqa}.

From the observational viewpoint,
this  complex scenario for the  lifetime of a \textbf{BH}-accretion disks system, opening  eventually   a new field of investigation in Astrophysics,  implies   a rich and diversified set of phenomena which may be associated with \textbf{RADs},
reinterpreting  the observations  analyzed so far in the single-torus framework,    in a new interpretive frame represented by the possibility of a multi-tori system.
Instabilities of
such configurations, we expect, may reveal of crucial significance for the High Energy Astrophysics related especially
to accretion onto supermassive \textbf{BHs}, and the extremely
energetic phenomena occurring in Quasars and \textbf{AGNs}
that could be observable by the planned X-ray observatory \texttt{\textbf{ATHENA}}\footnote{\url{http://the-athena-x-ray-observatory.eu/}}.
These configurations can be directly linked to the current models featuring the obscuration of galactic
Black Hole X-ray emission. The
radially oscillating tori of the couple could be related to
the high-frequency quasi periodic oscillations (\textbf{QPOs})
observed in non-thermal X-ray emission from compact
objects, keeping fingerprint of the discrete radial profile of the couple structure.
 Moreover, relatively indistinct excesses
 of the relativistically broadened  emission-line components were predicted, arising in a well-confined
radial distance in the accretion structure
  originating by a series of  episodic accretion events \citep{S11etal,KS10,Schee:2008fc,Schee:2013bya}.

 Here the \textbf{RADs} framework has been used  to investigate the influence of the magnetic field also in the formation of the single torus, as a limiting  case  of the \textbf{RADs} and hence in the formation of the multiple case too, comparing results of this study with the situation in absence of the  magnetic contribution. Differences between these two cases are particularly evident    in the unstable  phases due to the tori collision  and the accretion.
%according to the possibility of creating the toroidal magnetic field by kinetic dynamo effect as demonstrated.
 We shall  focus on the identification of a possible link between the \textbf{RAD} formation and features,  the \textbf{BH}  spin and  the relative rotation of the fluids in the \textbf{RAD},
looking for  a  correlation between two or more of these elements and the presence of a toroidal  magnetic field,  especially on emergencies of  the instability phases.
Particularly we analyze the situation for  a \emph{dual-accretion} phase when two tori are both accreting  onto Kerr attractors of a special class,  defined  through the \textbf{BH}  dimensionless spin  and determined by a special  relation  between the tori relative rotation. This special context reveals an interesting scenario in the  coupling between magnetic field  effects  and the  fluid rotation.
The choice of a purely azimuthal (toroidal)  magnetic field is  particularly adapted to the  disks symmetries considered here  and largely adopted  as initial setup for numerical  simulations  in  several general relativistic magnetohydrodynamic  (\textbf{GRMHD}) models sharing similar symmetries to the \textbf{RAD} considered here\citep{Luci}. From the methodological viewpoint, the magnetic field contribution has been  then  considered as part of the  exact general relativity effective potential  functions for  both  the fluid and \textbf{RADs}.
 Finally , we used  the  exact analytical magnetic field solution widely used and known as  the Komissarov solution\citep{Komissarov:2006nz}--and also  \citep{abrafra,Luci,EPL,adamek,Hamersky:2013cza,Karas:2014rka,[65],Slany:2013rml,Kovar:2011uh,Fragile:2017lbx,Gimeno-Soler:2017qmt} for applications in the context of accretion disks.

The structure of this article is  as follows:  In Section\il\ref{Sec:model} we  introduce the case of perfect fluid tori orbiting a central Kerr \textbf{BH}, and we set up the model for magnetized torus, discussing  the main quantities and notation used throughout this work.
 Section\il\ref{Sec:MRADa} contains the main results of our analysis, dealing  with the magnetized ringed accretion disk (\textbf{RAD}), by considering first the limiting case of non-magnetized \textbf{RAD} constituted by a couple of tori orbiting a central Kerr \textbf{BH}, and then  we concentrate our attention on the situation where a toroidal magnetic field  is for each component of the \textbf{RAD} system. This section closes with subsection\il\ref{Sec:q-less-1}, where some
 further considerations on the parameter choice follow and, by considering  an extended range of  variation for the magnetic field parameter, we discuss a very special class of \textbf{RADs} tori.
 {In Section\il\ref{Sec:notes} we add some further notes on the \textbf{RAD} instabilities considering also the phenomenological implications  and the influence of the toroidal magnetic field in the system stability.}
 Section\il\ref{Sec:Discussion} traces the conclusions of our investigation and we discuss  our results and observational consequences.

\section{Magnetized tori in  the Kerr spacetime}\label{Sec:model}
We consider toroidal  perfect fluids orbiting  in  the Kerr
spacetime background with metric tensor
\bea \label{alai} ds^2=-dt^2+\frac{\rho^2}{\Delta}dr^2+\rho^2
d\theta^2+\\\nonumber(r^2+a^2)\sin^2\theta
d\phi^2+\frac{2M}{\rho^2}r(dt-a\sin^2\theta d\phi)^2\ ,
\eea
in Boyer-Lindquist (BL)  coordinates
\( \{t,r,\theta ,\phi \}\).
Here $M$ is a mass parameter and the specific angular momentum is given as $a=J/M$, where $J$ is the
total angular momentum of the gravitational source and  $\rho^2\equiv r^2+a^2\cos\theta^2$, $\Delta\equiv r^2-2 M r+a^2$, in the following it will be also  convenient to introduce  the quantity
$\sigma \equiv\sin\theta$. We will consider  the Kerr Black Hole (\textbf{BH}) case defined by $a\in ]0,M[ $,  the extreme Black Hole source $a=M$, and the non-rotating  limiting case $a=0$, which is  the  Schwarzschild static metric.
 The horizons $r_-<r_+$ and the static limit $r_{\epsilon}^+$ are respectively
\bea
r_{\pm}\equiv M\pm\sqrt{M^2-a^2};\quad r_{\epsilon}^{+}\equiv M+\sqrt{M^2- a^2 \cos\theta^2},
%&&\nonumber
\eea
it is $r_+<r_{\epsilon}^+$ on the plane  $\theta\neq0$  and it is $r_{\epsilon}^+=2M$  on the equatorial plane $\theta=\pi/2$.
In the  {region $r\in]r_+,r_{\epsilon}^{+}$[} ({\em ergoregion}) it is  { $g_{tt}>0$} and $t$-Boyer-Lindquist coordinate becomes spacelike.
In this work we investigate toroidal  configurations of a perfect magnetized and non magnetized  fluids orbiting a Kerr attractor\citep{pugtot,abrafra,EPL,Pugliese:2012ub}.
Metric is independent of $\phi$ and $t$, as consequence of this  the covariant
components $p_{\phi}$ and $p_{t}$ of a particle four--momentum are
conserved along its  geodesic. Therefore, quantities\footnote{We adopt the
geometrical  units $c=1=G$ and  the $(-,+,+,+)$ signature, Latin indices run in $\{0,1,2,3\}$.  The   four-velocity  satisfy $u^a u_a=-1$. The radius $r$ has unit of
mass $[M]$, and the angular momentum  units of $[M]^2$, the velocities  $[u^t]=[u^r]=1$
and $[u^{\varphi}]=[u^{\theta}]=[M]^{-1}$ with $[u^{\varphi}/u^{t}]=[M]^{-1}$ and
$[u_{\varphi}/u_{t}]=[M]$. For the seek of convenience, we always consider the
dimensionless  energy and effective potential $[V_{eff}]=1$ and an angular momentum per
unit of mass $[L]/[M]=[M]$.}
\be\label{Eq:after}
{E} \equiv -g_{ab}\xi_{t}^{a} p^{b},\quad L \equiv
g_{ab}\xi_{\phi}^{a}p^{b}\ ,
\ee
are  constants of motion, where  $\xi_{t}=\partial_{t} $  is
the Killing field representing the stationarity of the Kerr geometry and  $\xi_{\phi}=\partial_{\phi} $
is the
rotational Killing field, the vector $\xi_{t}$ is   spacelike in the ergoregion.
%The
% momentum $p^a= \mu u^a$ of the  particle with  mass $\mu$ and four-velocity $u^{a}$
%can be normalized so that
%$g_{ab}u^{a}u^{b}=-k$, where $k=0,-1,1$ for null, spacelike and timelike
%curves, respectively.
In general,  we may interpret $E$, for
timelike geodesics, as representing the total energy of the test particle
 coming from radial infinity, as measured  by  a static observer at infinity, and  $L$ as the angular momentum  of the particle.
Furthermore,
Kerr metric \ref{alai} is invariant under the application of any two different transformations: $x^a\rightarrow-x^a$
  as one of the coordinates $(t,\phi)$ or the metric parameter $a$, and  the circular geodesic motion is  invariant under the mutual transformation of the parameters
$(a,L)\rightarrow(-a,-L)$. A consequence of this  we can  limit the  analysis of test particle circular motion  to the case of  positive values of $a$,
for corotating  $(L>0)$ and counterrotating   $(L<0)$ orbits.
Some  notable  radii regulate the particle dynamics, namely the \emph{marginally circular orbit} for timelike particles  $r_{\gamma}^{\pm}$,  the \emph{marginally  bounded orbit}  is $r_{mbo}^{\pm}$, and the \emph{marginally stable circular orbit} is $r_{mso}^{\pm}$ with angular momentum and energy  $(E_{\pm}. \mp L_{\pm})$ respectively,  where $(\pm)$ is for  counterrotating or corotating orbits with respect to the attractor\citep{Pu:Kerr,ergon,Pu:KN}.
In the case a non-magnetized tori  we may  consider a one-specie particle perfect  fluid (simple fluid),  where
\be\label{E:Tm}
T_{a b}=(\varrho +p) u_{a} u_{b}+\  p g_{a b},
\ee
is the fluid energy momentum tensor,  $\varrho$ and $p$ are  the total energy density and
pressure, respectively, as measured by an observer moving with the fluid. For the
symmetries of the problem, we always assume $\partial_t \mathbf{Q}=0$ and
$\partial_{\varphi} \mathbf{Q}=0$, being $\mathbf{Q}$ a generic spacetime tensor
(we can refer to  this assumption as the condition of  ideal hydrodynamics of
equilibrium).
  The timelike flow vector field  $u^a$  denotes now the fluid
four-velocity.
 We investigate in this work in particular  the case of a fluid circular configuration on  the fixed plane $\sigma=1$, defined by the constraint
$u^r=0$,  as for the circular test particle motion no
motion is assumed in the $\theta$ angular direction, which means $u^{\theta}=0$.
We assume moreover a barotropic equation of state $p=p(\varrho)$. While the  continuity equation %equation\il\ref)
% %
% \be
% \partial_{a}(\sqrt{-g} \rho U^a)=0
% \ee
 %
is  identically satisfied as consequence of the conditions.
 The Euler  equation for the pressure $p$ can be  written in the  non-magnetized case ($B=0$) as
\bea\label{Eq:scond-d}
\frac{\partial_{\mu}p}{\varrho+p}=-\frac{\partial}{\partial \mu}W+\frac{\Omega \partial_{\mu}\ell}{1-\Omega \ell},\\\nonumber W\equiv\ln V_{eff}(\ell),\quad \ell\equiv \frac{L}{E},\quad V_{eff}(\ell)=u_t
\eea
where $V_{eff}(\ell)$ is the \emph{effective potential}% and  the function $W$ is Paczynski-Wiita  (P-W) potential,
$\Omega$ is the relativistic angular velocity. Assuming  the fluid  is   characterized by the  specific  angular momentum  $\ell$  constant (see also \citet{Lei:2008ui}),  we consider  the equation for  $W:\;  \ln(V_{eff})=\rm{c}=\rm{constant}$ or $V_{eff}=K=$constant.
The procedure described in the present article
borrows from the  Boyer theory on the equipressure surfaces applied to a  thick  torus \citep{Boy:1965:PCPS:,abrafra}.
  The Boyer surfaces of the \textbf{RAD} tori are given by the surfaces of constant pressure  or\footnote{{More generally $\Sigma_{\mathbf{Q}}$ is the  surface $\mathbf{Q}=$constant for any quantity or set of quantities $\mathbf{Q}$.In this models the entropy is constant along the flow. According to the von Zeipel condition, the surfaces of constant angular velocity $\Omega$ and of constant specific angular momentum $\ell$ coincide \citep{M.A.Abramowicz,Chakrabarti0,Chakrabarti,Zanotti:2014haa} and  the rotation law $\ell=\ell(\Omega)$ is independent of the equation of state \citep{Lei:2008ui}.
}}  $\Sigma_{i}=$constant for \(i\in(p,\varrho, \ell, \Omega) \), where it is indeed $\Omega=\Omega(\ell)$ and $\Sigma_i=\Sigma_{j}$ for \({i, j}\in(p,\varrho, \ell, \Omega) \).

The  function  $V_{eff}(\ell)$  in equation\il\ref{Eq:scond-d}  is invariant under the mutual transformation of the parameters
$(a,\ell)\rightarrow(-a,-\ell)$, as for the case of test particle circular orbits we can limit our analysis to  positive values of $a>0$,
for corotating  $(\ell>0)$ and counterrotating   $(\ell<0)$ fluids.  More generally  we adopt the notation $(\pm)$  for  counterrotating or corotating matter  respectively.
In the  ringed accretion disks  system,  where a couple $(\cc_a, \cc_b)$ of  tori are orbiting  in   the equatorial plane of a central Kerr \textbf{BH}    with specific angular momentum $(\ell_a, \ell_b)$,   we need to introduce   the concept  of
 \emph{$\ell$corotating} tori,  defined by  the condition $\ell_{a}\ell_{b}>0$, and \emph{$\ell$counterrotating} tori  by the relations   $\ell_{a}\ell_{b}<0$,  the two $\ell$corotating tori  can be both corotating $\ell a>0$ or counterrotating  $\ell a<0$ with respect to the central attractor.

In the magnetized case,  following \citep{EPL,abrafra,Pugliese:2011aa}, we consider  an infinitely conductive plasma where $F_{ab}u^a=0$, and  $F_{ab}$ is the Faraday tensor, $u^a B_a=0$, where $B^a$ is the magnetic field and   $\partial_{\phi}B^a=0$ and $B^r=B^{\theta}=0$.
As noted in \citet{Komissarov:2006nz} the presence of a magnetic field with a relevant toroidal component can be   related  to the disk differential rotation, viewed as a generating  mechanism of the magnetic field,  for further discussion we refer  to \citep{Komissarov:2006nz,Montero:2007tc,Horak:2009iz,Parker:1955zz,Parker:1970xv,Y.I.I2003,R-ReS1999,Safarzadeh:2017mdy}, while we refer to   \citep{EPL,adamek,Hamersky:2013cza,Karas:2014rka,abrafra} where this solution is dealt  in detail in the context of accretion disks.
The Euler equation for this system has been exactly integrated for the background spacetime of Schwarzschild and Kerr \textbf{BHs}
 in  \citep{Komissarov:2006nz,Montero:2007tc,Horak:2009iz} with    a magnetic field is
\bea\label{RSC}
B^{\phi }=\sqrt{\frac{2 p_B}{g_{\phi \phi }+2 \ell  g_{t\phi}+\ell ^2g_{tt}}}\quad\mbox{or alternatively}\\\nonumber    B^{\phi }=\sqrt{{2 \mathcal{M} \omega^q}} \left(g_{t \phi }g_{t \phi }-g_{{tt}}g_{\phi \phi }\right){}^{(q-2)/2} V_{eff}(\ell)
\eea
where
\(
p_B=\mathcal{M} \left(g_{t \phi }g_{t \phi }-g_{{tt}}g_{\phi \phi }\right){}^{q-1}\omega^q
\) is the magnetic pressure,
$\omega$ is the fluid enthalpy, $q$  and $\Mie$ are constant; we assume moreover a barotropic equation of state.  Equation\il\ref{Eq:scond-d}  has been used in
 second term of  equation\il\ref{RSC}.
According to our set-up we  introduce a deformed (magnetized) \emph{ Paczy{\'n}ski potential function} and the
  Euler  equation  \ref{Eq:scond-d} becomes:
{\small{
\bea\label{Eq:Kerr-case}
&&\partial_{\mu}\tilde{W}=\partial_{\mu}\left[\ln V_{eff}+ \mathcal{G}\right]\, \mbox{where}
%\\\nonumber
%&&(a=0):\left.\mathcal{G}(r,\theta)\right|_{a=0}= \Sie \left[(r-2) r \sigma^2\right]^{(q-1)};
\\\nonumber
&&(a\neq0):\mathcal{G}(r,\theta)=\Sie \left(\mathcal{A} V_{eff}^2\right)^{q-1}=\Sie\left(g_{{t\phi }} g_{t\phi}-g_{tt} g_{\phi \phi}\right)^{q-1};\\ &&
\nonumber\mbox{and}\;\mathcal{A}\equiv\ell ^2 g_{tt}+2 \ell  g_{t\phi}+g_{\phi \phi },\, \Sie\equiv\frac{q \mathcal{M} \omega ^{q-1}}{q-1}
\eea}}
  parameter $\Sie$ is sketched in  Figure\il\ref{Fig:SaidSynergy}.
We therefore consider  the equation for the
\(
\tilde{W}\equiv \mathcal{G}(r,\theta)+\ln(V_{eff})=K
\).   The toroidal surfaces  are obtained from the equipotential surfaces \citep{Boy:1965:PCPS:,EPL},
where there is
\bea\label{Eq:goood-da}
&&\widetilde{V}_{eff}^2\equiv V_{eff}^2 e^{2 \Sie \left(\mathcal{A} V_{eff}^2\right){}^{q-1}}=
\\
\nonumber&&\frac{\left(g_{t \phi} g_{t \phi}-g_{tt} g_{\phi \phi }\right) \exp \left(2 S \left(g_{t \phi} g_{t \phi}-g_{tt} g_{\phi \phi }\right)^{q-1}\right)}{\ell ^2 g_{tt}+2 \ell  g_{t \phi}+g_{\phi \phi }}=K^2.
\eea
Potential $\widetilde{V}_{eff}^2$, for $\Sa=0$ reduces to  the effective potential ${V}_{eff}^2$ for the non-magnetized case in
equation\il\ref{Eq:scond-d}.
The equipressure surfaces, $K=$constant,   could be closed, $\cc$, determining equilibrium configurations, or open $\oo_{\times}$ (related to  ``proto-jet'' configurations\citep{open}). The special case of cusped $\cc_{\times}$ equipotential surfaces allows for the accretion onto the central Black Hole, due to the   Paczynski-Wiita  (P-W)  hydro-gravitational instability  mechanism occurring at the cusp $r_{\times}$, \citep{Pac-Wii}: the outflow of matter through the cusp occurs due to an instability in the balance of the gravitational and inertial forces and the pressure gradients in the fluid, i.e., a mechanism of violation of mechanical equilibrium of the tori--Figure\il\ref{Griegn}.
For each torus, the extrema of the effective potential functions fix   the center $r_{\odot}$,  as minimum point $r_{\min}$ of the effective potential and the maximum point for the hydrostatic pressure. The inner edge $r_{\times}$ of the  accreting  torus, when accretion occurs,  corresponds to the maximum  point $r_{\max}$  of  the effective potential, also the minimum point for the hydrostatic pressure. The inner and outer edges  of an equilibrium torus are also strongly constrained.
 The inner edge  of the Boyer surface  is at $r_{in}\in[r_{\max},r_{\min}[$  on the equatorial plane,
while the outer edge  is at $r_{out}>r_{\min}$  on the equatorial plane. For a  discussion on the definition and location of the inner edge of the accreting torus see \citep{Krolik:2002ae,BMP98,2010A&A...521A..15A,Agol:1999dn,Paczynski:2000tz}.

In the following,  for any quantity $\mathbf{Q}$ and radius $r_{\bullet}$ we adopt the notation $\mathbf{Q}_{\bullet}\equiv \mathbf{Q}(r_{\bullet})$,  for example  there is $\ell_{\mso}^+\equiv\ell_+(r_{\mso}^+)$.
Then
1. for fluid specific angular momentum $\ell$ in
$\mp\ell^{\pm}\in\mp L^{\pm}_1\equiv[\mp \ell_{\mso}^{\pm},\mp\ell_{\mbo}^{\pm}[$     topologies $(\cc_1, \cc_{\times})$ are possible, the $\cc_1$ indicated a non-accreting topology  $\cc$ with specific angular momentum $\ell\in L_1$
where  $r^{\pm}_{\times}\in]r^{\pm}_{\mbo},r^{\pm}_{\mso}]$ %and center of maximum pressure  $r_{\odot}\in]r_{\mso},\bar{\mathfrak{r}}_{(\mbo)}]$;
2. for
$\mp\ell^{\pm}\in\mp L^{\pm}_2\equiv\mp \ell^{\pm}\in[\mp \ell_{\mbo}^{\pm},\mp\ell_{\gamma}^{\pm}[ $    topologies    $(\cc_2, \oo_{\times})$ are possible,  with unstable point  $r^{\pm}_{j}\in]r^{\pm}_{\gamma},r^{\pm}_{\mbo}]$.   %and  center of maximum pressure $r_{\odot}\in]\bar{\mathfrak{r}}_{(\mbo)},\bar{\mathfrak{r}}_{(\gamma)}]$;
3. for
$\mp\ell^{\pm}\in\mp L^{\pm}_3\equiv\mp \ell^{\pm}\equiv\ \ell \geq\mp\ell_{\gamma}^{\pm} $     only equilibrium torus  $\cc_3$ is possible.

Similarly to the non magnetized case (where the effective potential is $V_{eff}(r;\ell, a))$, the function $\widetilde{V}_{eff}(r;\ell, a,\Sa, q)$  defined in equation\il\ref{Eq:goood-da}  may be regarded as   an effective potential function encoding the centrifugal and curvature binding effects of the spacetime  together with the magnetic pressure force ($q$), essentially regulated by the $\Sa$ values.
Therefore it is important to discuss the range of  variation for the $(q, \Sa)$ couple.
Section\il\ref{Sec:q-less-1}  addresses {further considerations on the parameter choice.}
We note that a negative solution for $\Sa$ may appear  for $q<1$;  we shall briefly consider also the case of tori in this more general situation  in section\il\ref{Sec:q-less-1}.

Here we note that, in the limiting case $q=0$,
the magnetic field $B$, does not depend on the fluid enthalpy, furthermore  equation\il\ref{Eq:goood-da} for  $q=0$  is $V_{eff}={\rm{K}}$, this means that the magnetic field $B^{\phi}\left|_{q=0}\right.$ does not effect the Boyer surfaces.

It is therefore worth to consider some limits, with  the coefficient $ \widetilde{V}_{eff}^2[n]$ of $\Sie^n$ in the expansion of $\widetilde{V}_{eff}^2$ around  $\Sie=0$:
\bea\nonumber
&&
\mbox{for}\, \mathcal{S}\approx 0\,
 \widetilde{V}_{eff}^2[n]= \frac{2^n}{n!} V_{eff}^2 \left(\mathcal{A} V_{eff}^2\right)^{n (q-1)}\,  n\geq 0,
  \\\label{Eq:dyna-il}
  &&
  \mbox{thus}\,   \widetilde{V}_{eff}^2= V_{eff}^2+\frac{2 \mathcal{S} \left(\mathcal{A} V_{eff}^2\right){}^q}{\mathcal{A}}+\mathbf{\mathrm{O}}\left(\Sie^2\right)
 \eea
where $\mathbf{\mathrm{O}}(\mathbf{Q}^\kappa)$ is for terms of the order greater of equal then $\mathbf{Q}^{\kappa}$ for any quantity $\mathbf{Q}$.
As $\Sie=\Sie(q)$ we   consider therefore the  coefficient $\mathcal{\Sie}_n$ of $(q-1)^n$ in the expansion of $\mathcal{\Sie}$ around  $q=1$:
\bea\label{Eq:prod-gap}
\mathcal{\Sie}_n=\frac{\mathcal{M} \ln ^n(\omega ) (n+\ln (\omega )+1)}{\Gamma (n+2)}\,\mbox{for}\, n\geq0\,\mbox{ and }\, q\gtrapprox1,
\eea
where $\Gamma(x)$ is the Euler gamma function--Figures\il\ref{Fig:SaidSynergy}.
\begin{figure}
\centering
\includegraphics[width=.81\columnwidth]{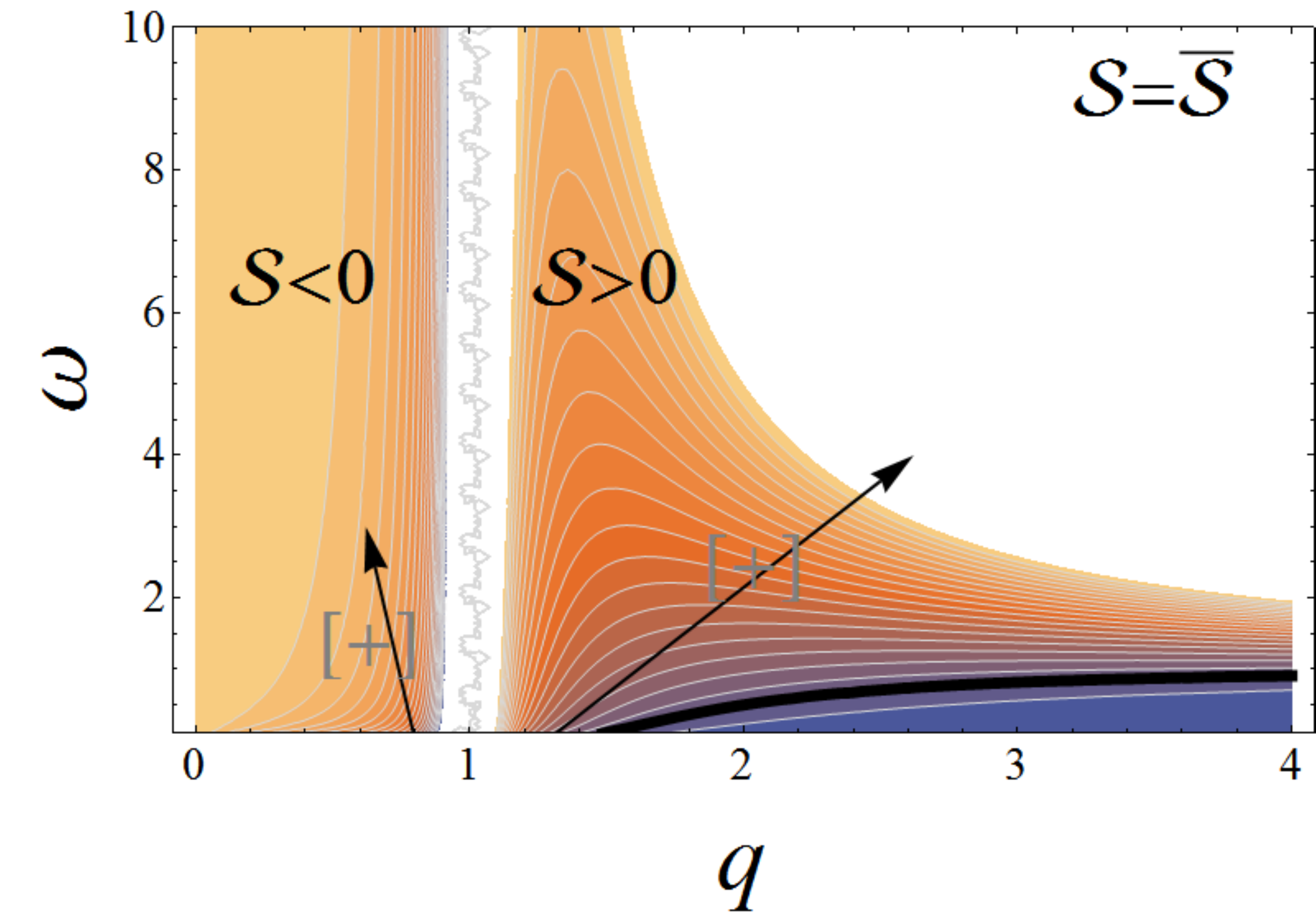}\\
\includegraphics[width=.81\columnwidth]{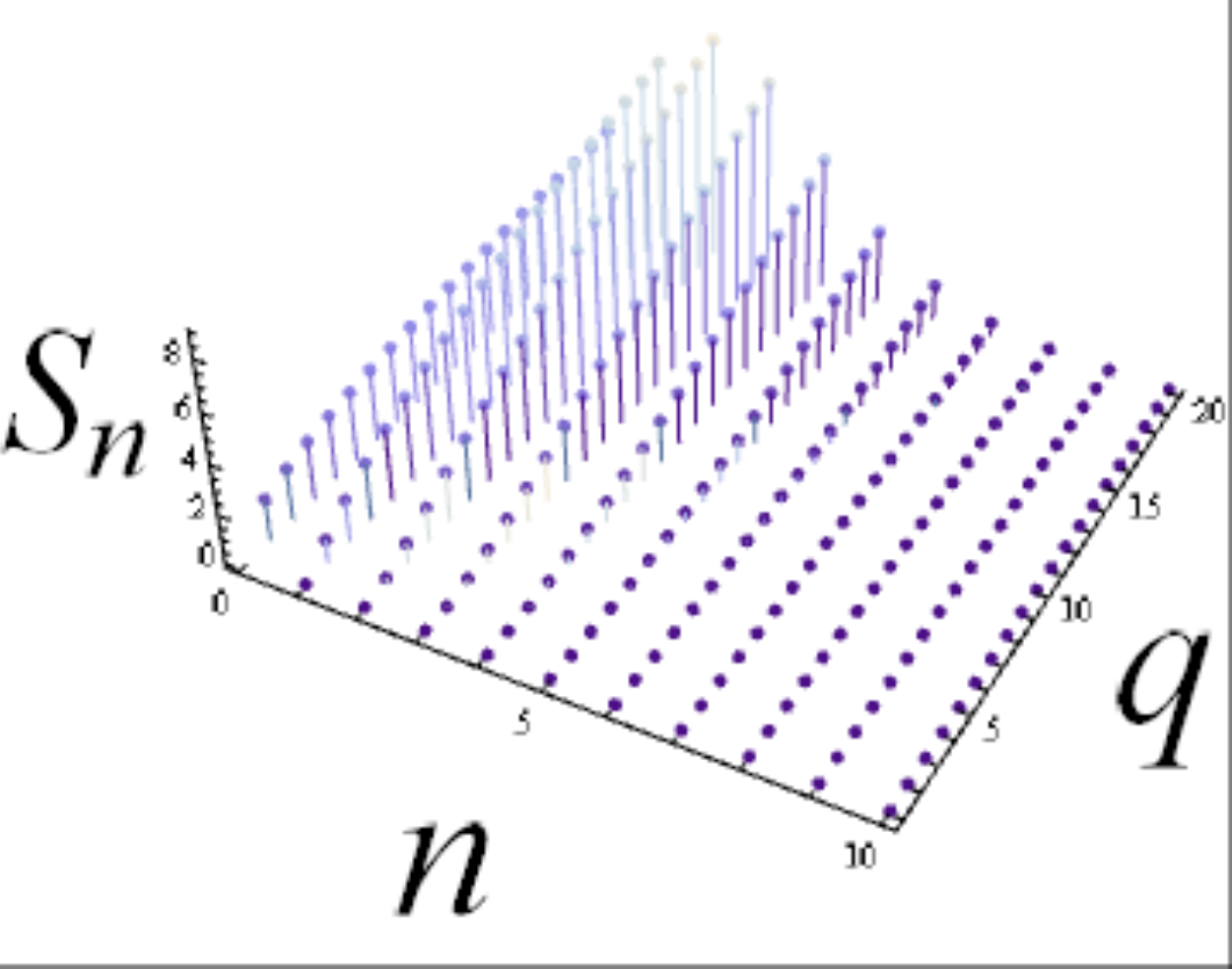}
%\\
%\includegraphics[width=0.4\hsize,clip]{Sometimes}
%\includegraphics[width=0.4\hsize,clip]{Edbott}
\caption[font={footnotesize,it}]{Upper panel: Constant values of  $\Sa=\bar{\Sa}$, introduced  in equation\il\ref{Eq:Kerr-case}  as function of  the fluid enthalpy $\omega$ and magnetic parameter $q$ (dimensionless quantities are considered), regions of positive $\Sa>0$ and negative values $\Sa<0$ are considered. Black thick line is  $\Sa=0$, arrows mark the increasing values of $\Sa$ parameters.  Bottom panel:  coefficient $\mathcal{\Sie}_n$ of $(q-1)^n$ in the expansion of $\mathcal{\Sie}$ around  $q=1$--see equation\il\ref{Eq:prod-gap}, the situation for   $q\approx 1$ and $r\rightarrow \infty$, and  the limits  $\Sa=1$  and $\Sa=0$ are shown. }
\label{Fig:SaidSynergy}
\end{figure}
In Section\il\ref{Sec:MRADa}  we  consider in details the case of two magnetized tori orbiting a Kerr central \textbf{BH} focusing first  on  the limiting case   of non-magnetized \textbf{RAD} system ($\Sa=0$) of the order two, made up by two orbiting tori.

\section{Magnetized ringed accretion disks}\label{Sec:MRADa}

We start by  solving the equation for the critical points of the function  $\tilde{V}_{eff}$
with respect to the fluid  specific angular momentum  obtaining, similarly to the non-magnetized  case the fluid specific angular momentum  ${\ell}^{\pm}(r)$ replaced by  the solution  $\widetilde{\ell}^{\pm}(r):\, \partial_r\tilde{V}_{eff}=0$, for counterrotating and corotating magnetized  fluids  respectively

\begin{strip}
{{
\bea\label{Eq:dilde-f}
\nonumber
&&\widetilde{\ell}^{\mp}\equiv\frac{\Delta \left(a^3+a r \left[4 \Qa (r-M) \Sie \Delta^{\Qa}+3 r-4\right]\mp\sqrt{r^3 \left[\Delta ^2+4 \Qa^2 (r-1)^2 r \Sie^2 \Delta ^{2 \Qa+1}+2 \Qa (r-1)^2 r \Sie \Delta ^{\Qa+1}\right]}\right)
}{
%\\\nonumber&&
a^4-a^2 (r-3) (r-2) r-(r-2) r \left[2 \Qa (r-1) \Sie \Delta ^{\Qa+1}+(r-2)^2 r\right]}
\\&&\label{Eq:polis-ll}
\mbox{where there is }\, \lim_{\mathcal{\Sie}\rightarrow0}\widetilde{\ell}^{\mp}=\lim_{q\rightarrow 1}\widetilde{\ell}^{\mp}=\ell^{\pm},
\eea}}
\end{strip}
 dimensionless quantities  $r\rightarrow r/M$ and $a\rightarrow a/M$ have been used--see Figures\il\ref{Fig:colin} and Figures\il\ref{Fig:SaidSynergymat}.
\begin{figure}
\includegraphics[width=1\columnwidth,clip]{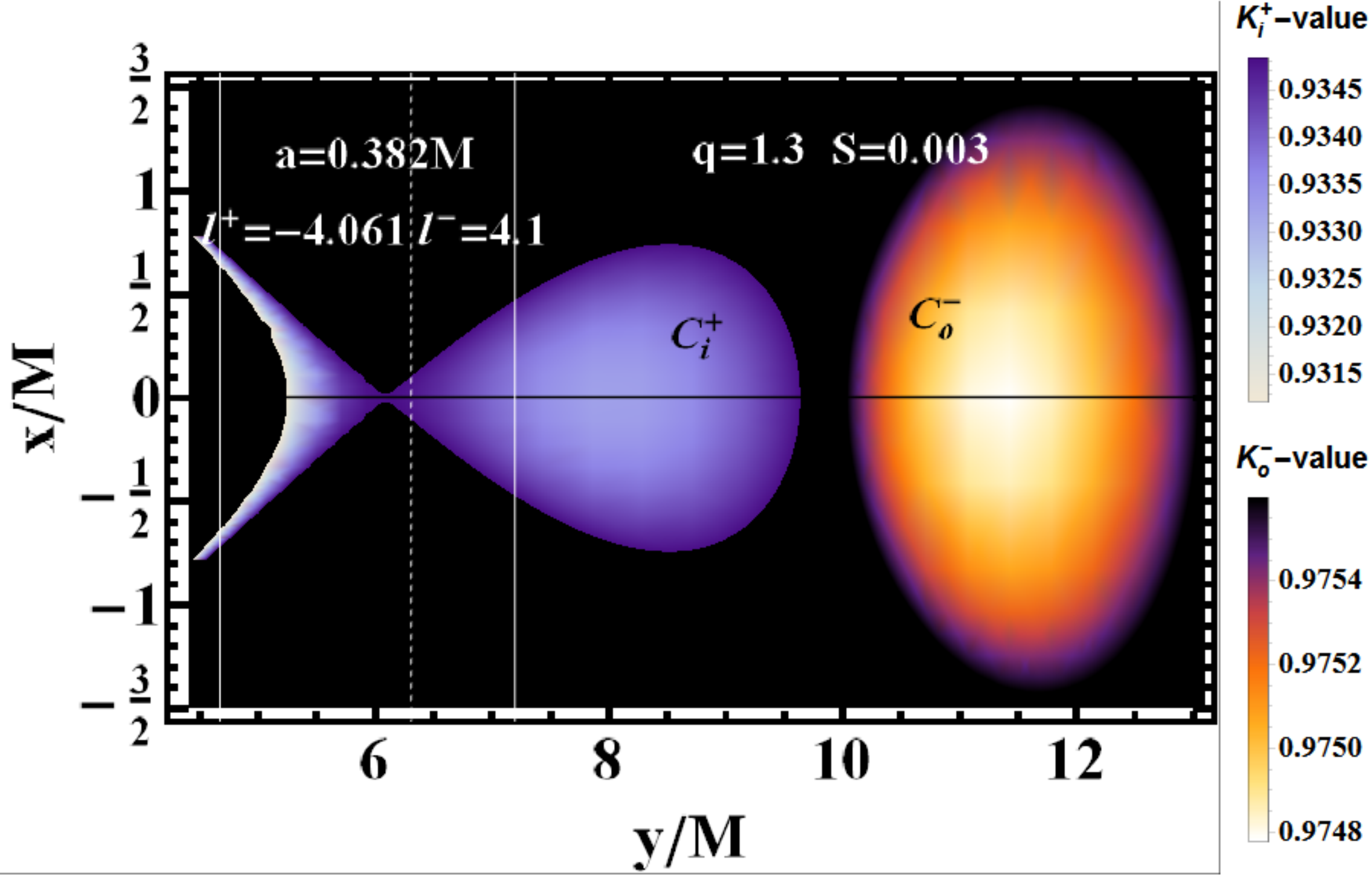}\\
\includegraphics[width=1\columnwidth,clip]{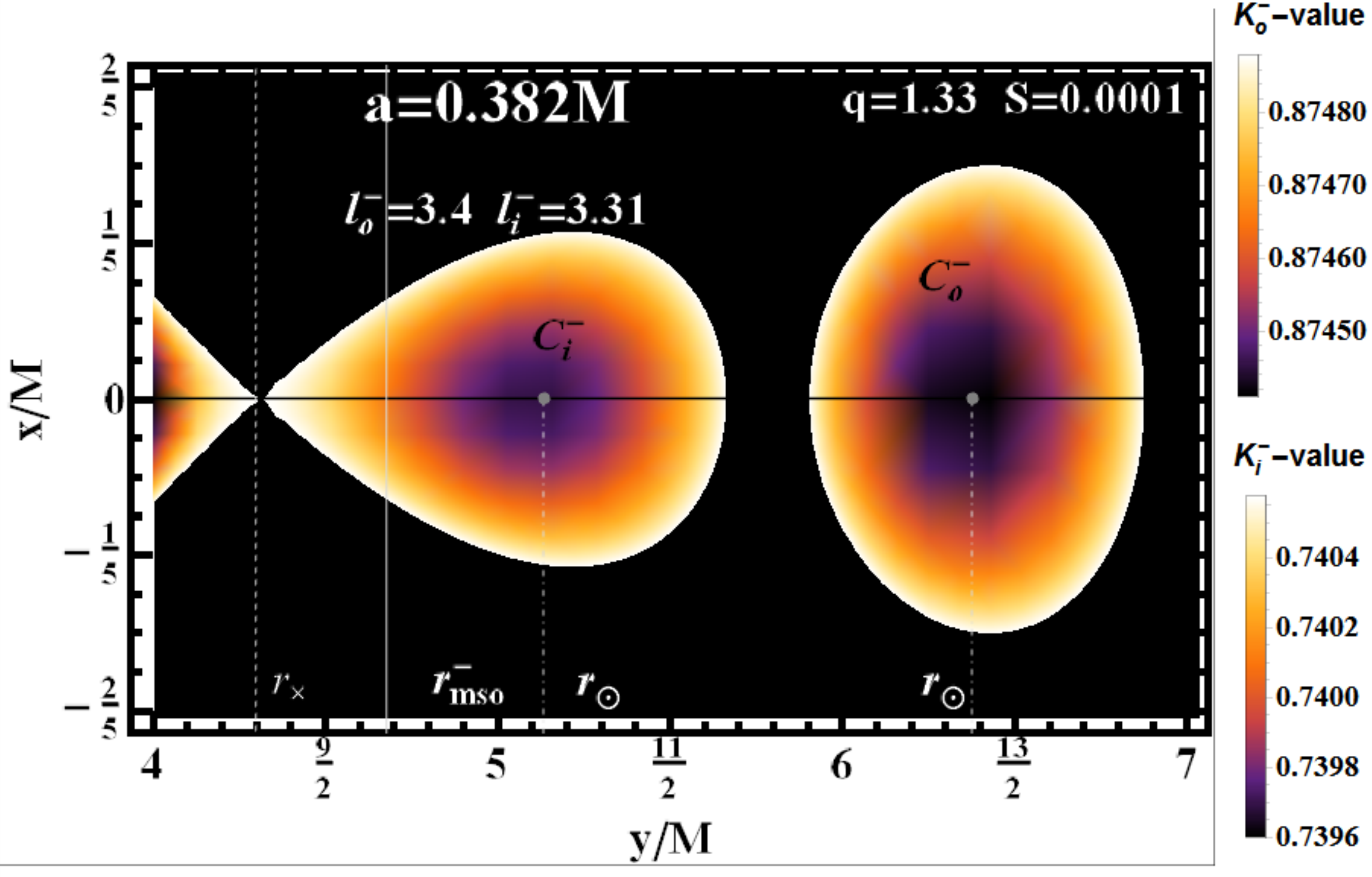}\\
\includegraphics[width=1\columnwidth,clip]{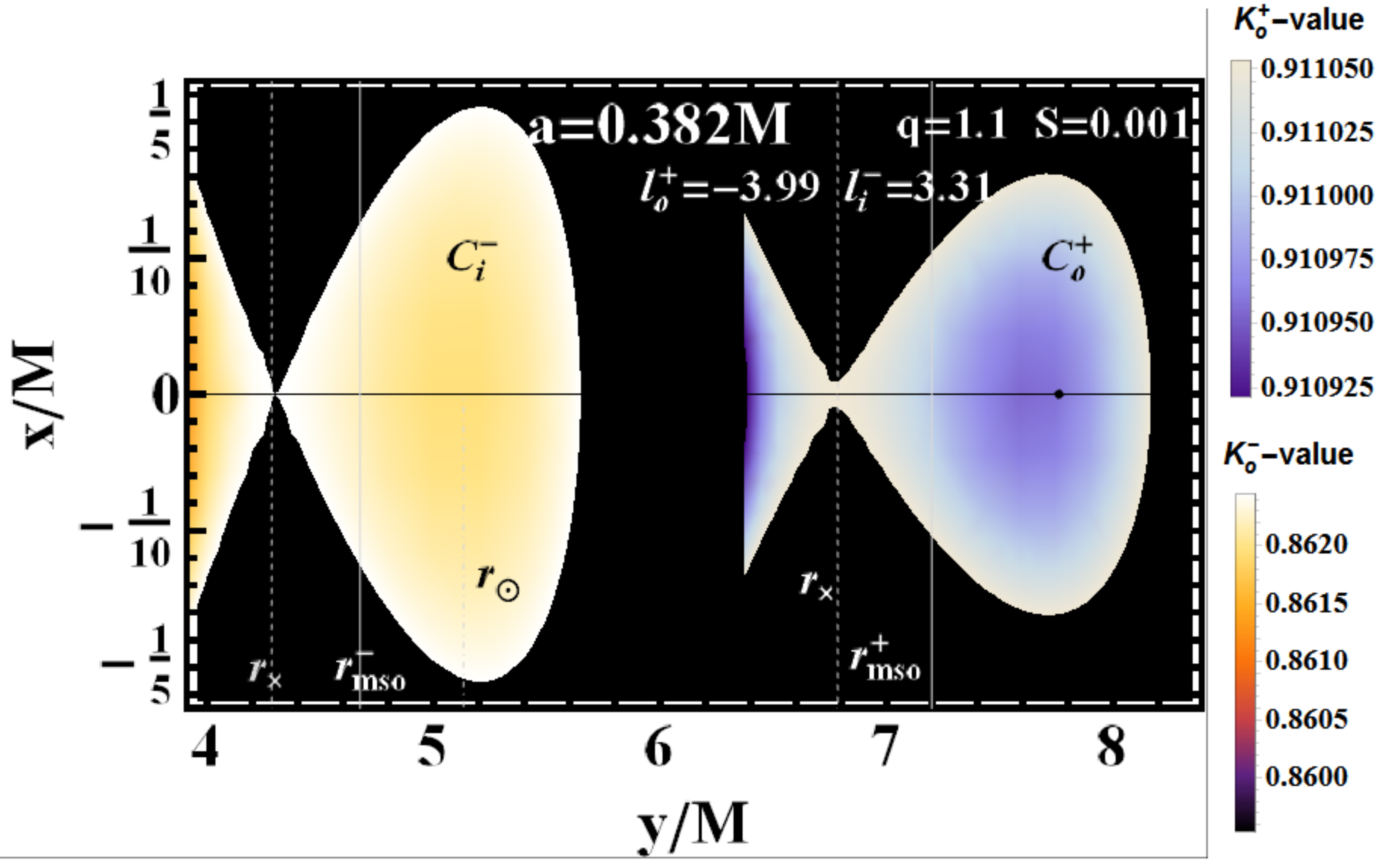}%\\
\caption{Density plots.  \emph{Upper-first} panel: $\cc^+_{\times}<\cc^-$ \textbf{RAD}; \emph{Upper-second} panel:
$ \cc^-_{\times}<\cc^-$ \textbf{RAD}.
\emph{Bottom-}panel: \textbf{RAD}
 $ \cc_{\times}^-<\cc_{\times}^+$.  $(x, y)$ are Cartesian coordinates. Magnetic parameters $(q,\Sa)$, BH spin $a/M$ and fluids specific angular momentum are signed $(\ell)$.}
\label{Griegn}
\end{figure}
We  obtained a specific fluid angular  momentum expression which explicitly includes the dependence of the field through the  $\Sa$ and  $q$ parameters.
 Then, we note also   in equation\il\ref{Eq:dilde-f} the explicit dependence of  $\widetilde{\ell}_{\pm}$ on the parameter $\Qa=q-1$--Figures\il\ref{Fig:colin} and \ref{Fig:SaidSynergymat}.
 Limits \ref{Eq:polis-ll}, furthermore, are consistent with the analysis in equations\il\ref{Eq:dyna-il}--\ref{Eq:prod-gap} for  the asymptotic behavior in the same regions of the parameter space.
 Therefore we can address the   comparison with the   non-magnetized case by considering the parameters  $\Sa=0$ or $q=1$.
 Nevertheless before to consider the effects of toroidal magnetic field  it is appropriate to further comment  the situation for the non-magnetized  \textbf{RADs}  \citep{ringed,open,dsystem}.

 In the following  we consider  specific tori couples or \emph{seeds}. As specified in \citet{open},  the study of these configurations allows both the direct characterization of the system, consisting of only
two accretion disks around a central attractor,  i.e. a \textbf{RAD} of the order $n=2$,  and it also simplifies  the analysis of  the more general case of multiple  toroidal configurations orbiting a single central  attractor.
The study of \textbf{RADs}  made by more then two orbiting  toroidal  configurations could  be carried out considering composition of   \emph{seed}   tori couples.
Therefore we can concentrate our attention here on two tori with parameters   $(\ell_i, \ell_o)$ and $(K_i, K_o)$, for  the inner ad outer tori respectively with respect to the central \textbf{BH} say, introducing notation $\lessgtr$, there is   $\cc_i<\cc_o$ for the relative location of the configurations.
The analysis of multiple toroidal disks can be then   further simplified by considering appropriate boundary conditions on  a properly defined  ``\textbf{RAD} effective potential'' which,  for a seed, may be defined as follows for a \textbf{RADs} { of the order $n=2$}:
{\small{ %
\bea\label{Eq:def-partialeK}
&&\left.\widetilde{V}_{eff}^{\mathbf{\cc}^2}\right|_{K}\equiv \widetilde{V}_{eff}^{i}\Theta(-K_i)\bigcup \widetilde{V}_{eff}^{o}\Theta(-K_o),\,\mbox{alternately}
\\&&
\nonumber\label{Eq:Vcomplessibo}
\widetilde{V}_{eff}^{\mathbf{\cc}^2}\equiv \widetilde{V}_{eff}^{i}(\ell_i)\Theta(r_{\odot}^{o}-r)\Theta(r-r_+) \widetilde{V}_{eff}^{o}(\ell_o)\Theta(r-r_{\odot}^{i}),
\eea}}
where $\Theta$ is  the Heaviside (step) function  such that  for example $\Theta(-K_i)=1$  for $\widetilde{V}_{eff}^{i}<K_i$ and $\Theta(-K_i)=0$ for $\widetilde{V}_{eff}^{i}>K_i$.
Note that we adopt the notation $\tilde{\ell}$, for the specific  angular momentum  in the magnetized case mainly   when it is  regarded as function of $(r;a, \Sa,q)$--equation\il\ref{Eq:dilde-f}; On the other hand, as  when the specific fluid angular momentum is considered as a parameter, for easy of reference, we  use simplified notation $\ell$,  when not otherwise specified--equation\il\ref{Eq:Vcomplessibo}.

 \medskip

 \textbf{Preliminary notes on the  RADs and the non-magnetized case}

\medskip

In    the \emph{non-magnetized case},
 accreting \textbf{RADs} couples may turn in the following four  cases \emph{only}:
{\textbf{(a)}} $\cc_{\times}^{\pm}< \cc^{\pm}$, {\textbf{(b)}} $\cc_{\times}^{+}< \cc^{\pm}$, {\textbf{(c)}}
$\cc_{\times}^{-}< \cc^{\pm}$ and {\textbf{(d)}} $\cc_{\times}^{-}< \cc_{\times}^{+}
$.
In the case {\textbf{(a)}}, describing $\ell$corotating tori or any couple around a static ($a=0$) attractor,    only the inner torus of  the couple is   accreting onto the central Black Hole.
The \textbf{RADs} with an  $\ell$counterrotating   couples, distinguish three major  classes of \textbf{BH} attractors defined by spin ranges with boundaries  in geometries characterized by spin
$a_{1}\equiv0.4740M $,  $ a_{2}=0.461854M$  and $a_{3}\equiv0.73688 M$ \footnote{The origin of these special spins can be retraced in the geometric properties of the  Kerr spacetime and the fluid dynamics,   quite independently by the  rotational law (specific angular momentum definition)--for more discussion see \citep{dsystem,letter,Lei:2008ui}. %} $a_1:\;r_{(\mbo)}^+=r_{({\mathrm{\gamma}}}^-$,\quad
% :$a_2\;r_{(\mbo)}^-=r_{\mso}^+,
%$    $a_3\;r_{({\mathrm{\gamma}}}^-=r_{\mso}^+,
}.
Couples
$\cc_{\times}^{\pm}<\cc^{\pm}$, \textbf{(a)}, and  $\cc_{\times}^-<\cc^{+}$, \textbf{(c)}, may form in  \emph{all} spacetimes where  $a\in[0, M]$ \citep{dsystem}. % (this can be easly inferred also from the  geodesic structure of the Kerr spacetime).
{On the other  hand, a   $\cc_{\times}^{-}< \cc_{\times}^{+}$ couple which features   a double accretion\footnote{We stress  that this special seed is particularly interesting, noting  that any \textbf{RADs} is to be  considered as a geometrically thin accretion disk \citep{ringed}. Then  only for this  special couple a ``screening''  effect may occure with corotating non-accretion disk between the two accreting tori of the \textbf{RAD} \citep{Marchesi,Gilli:2006zi,Marchesi:2017did,Masini:2016yhl,DeGraf:2014hna,Storchi-Bergmann}.  This mechanim envisages therefore a special ``inter-disk'' activity with greater potentiality also in view  of a possible jet-accretion correlation--see also \citet{dsystem,letter,long} and \citep{KJA78,AJS78,Sadowski:2015jaa,Lasota:2015bii,Lyutikov(2009),Madau(1988),Sikora(1981)}},  \textbf{(d)}-case,  can be observered in all  Kerr geometries $a\neq0$,  but  the slower is  the  \textbf{BH} ($a\lessapprox a_{1}$) the  lower must be the specific angular momentum $\ell_-$ of  the inner corotating torus and the smaller is the tori spacings.
Finally,  couples $\pp^+<\cc^-$,  where $\pp$ stays  for an accreting $(\cc_{\times})$  or non-accreting $(\cc)$ torus,  \textbf{(b)}, can be observable in  any spacetime $a\in[0,M]$, although \emph{only} around Kerr attractor with
   $a\in[0,a_{2}[$  the corotating, non-accreting, torus  $\cc^-$   approaches  the instability  ($r_{\times}\gtrapprox r_{\mso}^-$) in the \textbf{RAD} seed.
  Morever, the  {faster}  is the Kerr attractor ($a\gtrapprox a_{3}$),   the {farther away} {($r_{\odot}>r_{({\mathrm{\gamma}})}^-)$  should be the outer torus to prevent collision {\citep{dsystem,open,letter}}. A torus  screening effect in this case  can occur only with  coroting inner screening non-accreting disks. In fact,
an accreting   \emph{corotating} torus {must  be}  the inner one of the couple while the outer counterrotating torus can be non-accreting or in accretion.}
If there is a  $\cc_{\times}^-$ torus, or   if the attractor is {static},  then  no  inner (corotating or counterrotating) torus {can} exist, and  then  $\cc_{\times}^-$ is part of $\cc_{\times}^-<\cc^-$  couple or of a $\cc_{\times}^-<\pp^+$  one.
A  corotating torus can be the outer  of a couple of tori with an   inner counterrotating accreting torus. Then the outer torus may be corotating (non  accreting), or counterrotating in accretion or  {non-accreting}.    Both
the inner corotating  and the outer counterrotating  torus of the couple  can accrete onto the attractor.
A counterrotating  torus can therefore reach the instability being the inner one of  any  couple, or the outer torus of an $\ell$counterrotating  couple.
Then it is worth noting that if  the \emph{accreting} torus is \emph{counterrotating} with respect to the Kerr attractor, i.e.  a $\cc_{\times}^+$, then  there is \emph{no} inner  counterrotating torus, but a couple may be formed as a   $\cc_{\times}^+<\cc^{\pm}$ or as a $\pp^-<\cc_{\times}^+$ one.
  %%%%%%%%%%%%%%%%%%%%%%%%%%%%%%%%%%%%%%%%%%%%%%%%%%%%%%%%%%%
%
%%
 We propose here the analysis of the four cases for  the  magnetized fluids by  directly integrating the Euler equations,   as in  Figures\il\ref{Griegn}, and using proper model constraints  on the effective potential \ref{Eq:def-partialeK}.

 \medskip

 \textbf{The magnetized case}

 \medskip
 It is convenient  to take a  closer look at  the relation between   $\Sa$ and $q$.
 The different dependence  of $\tilde{\ell}^+$ and $\tilde{\ell}^-$, on the parameter couple $(\Sa,q)$, is enlighten in Figures\il\ref{Fig:colin} and \ref{Fig:SaidSynergymat}.
\begin{figure}
\centering
\includegraphics[width=.91\hsize,clip]{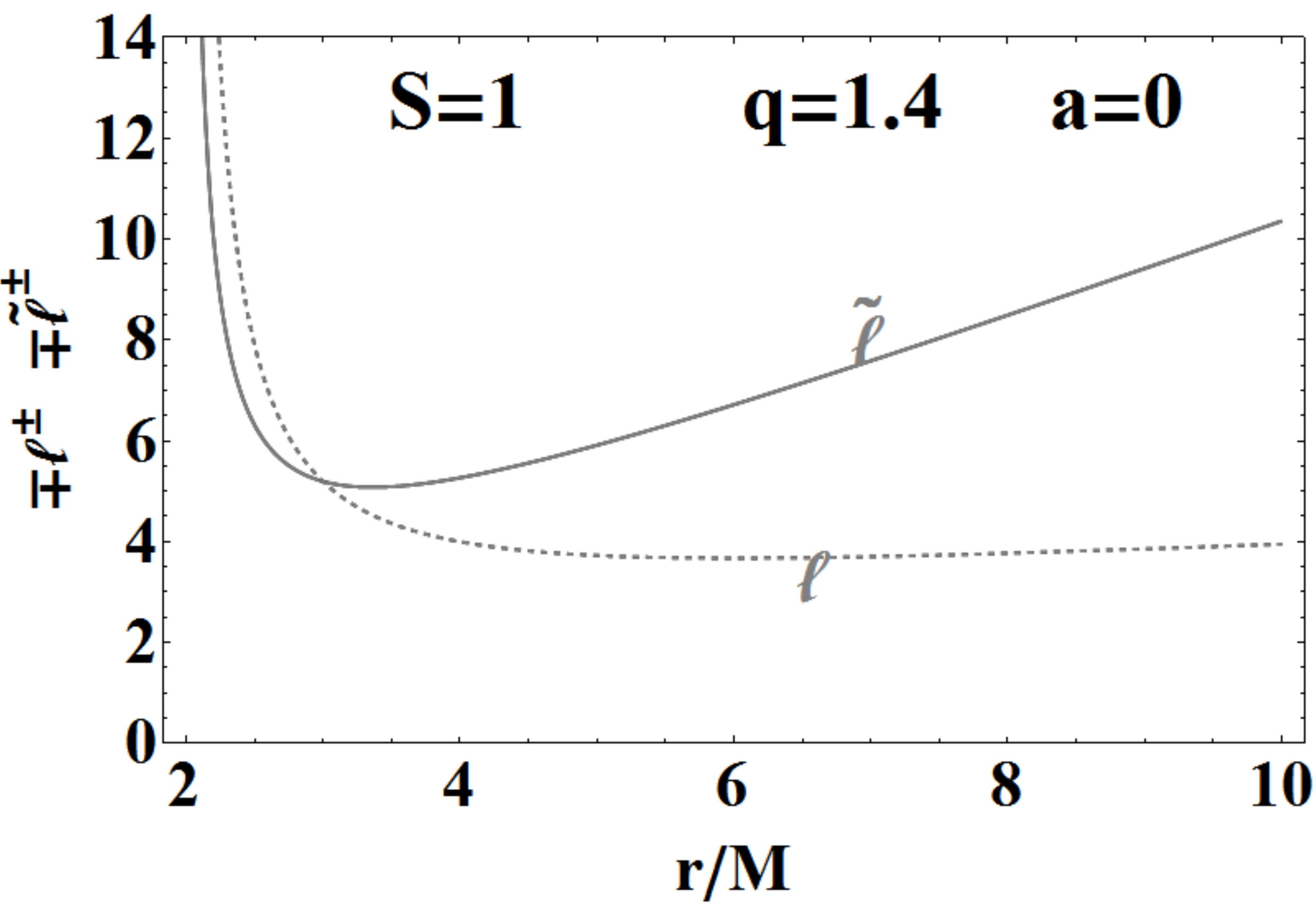}\\
\includegraphics[width=.91\hsize,clip]{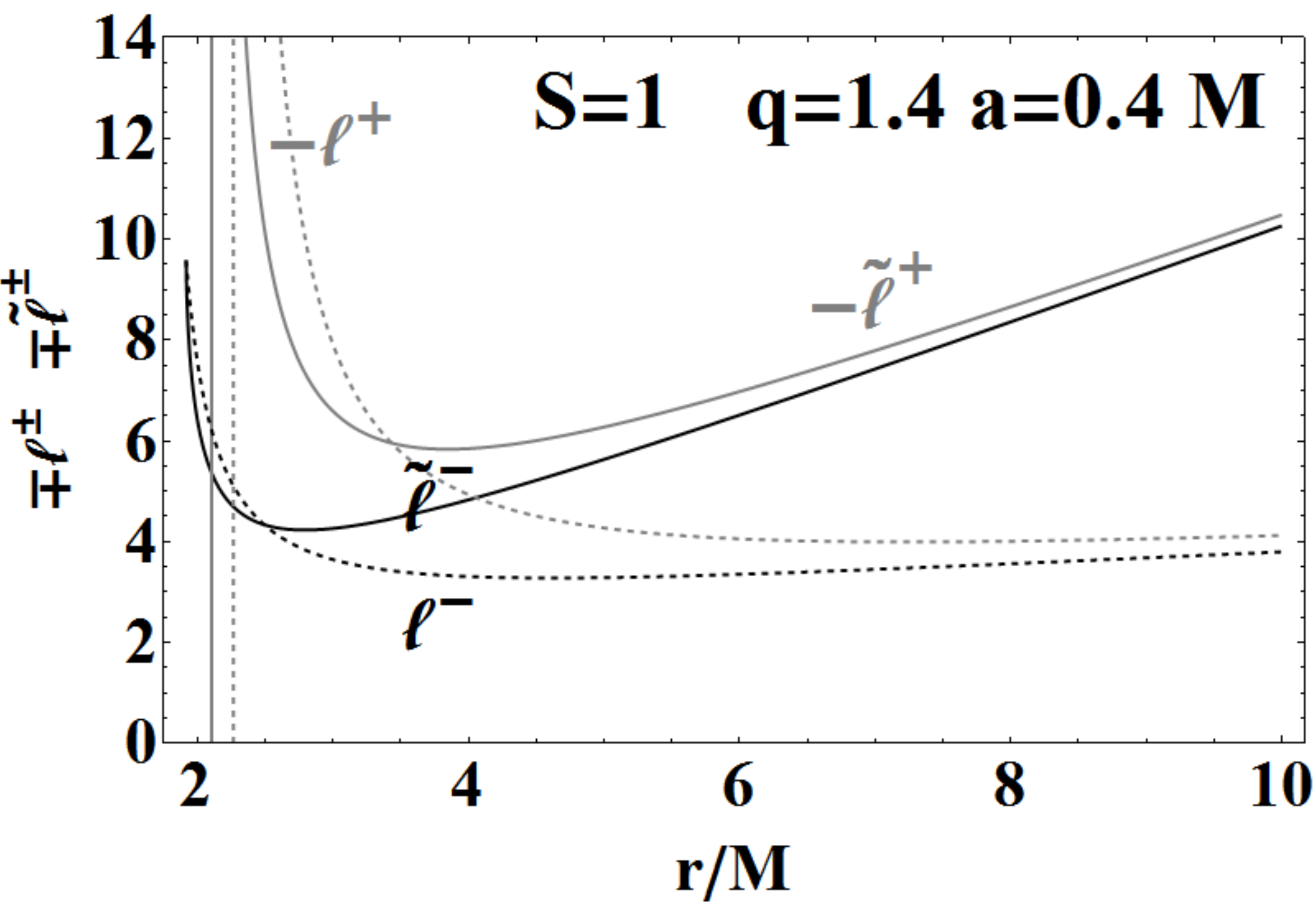}\\
\includegraphics[width=.91\hsize,clip]{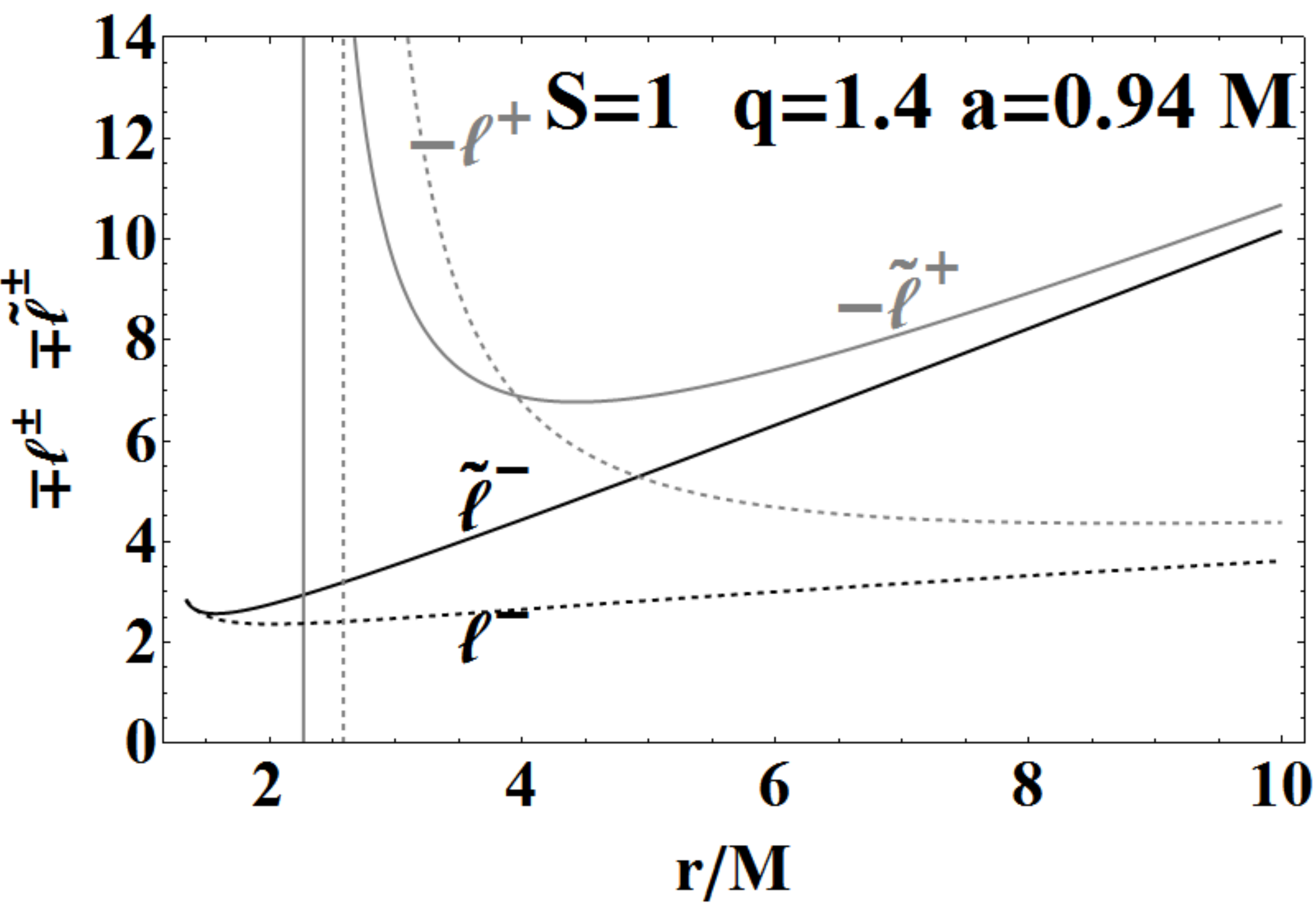}\\
\includegraphics[width=.91\hsize,clip]{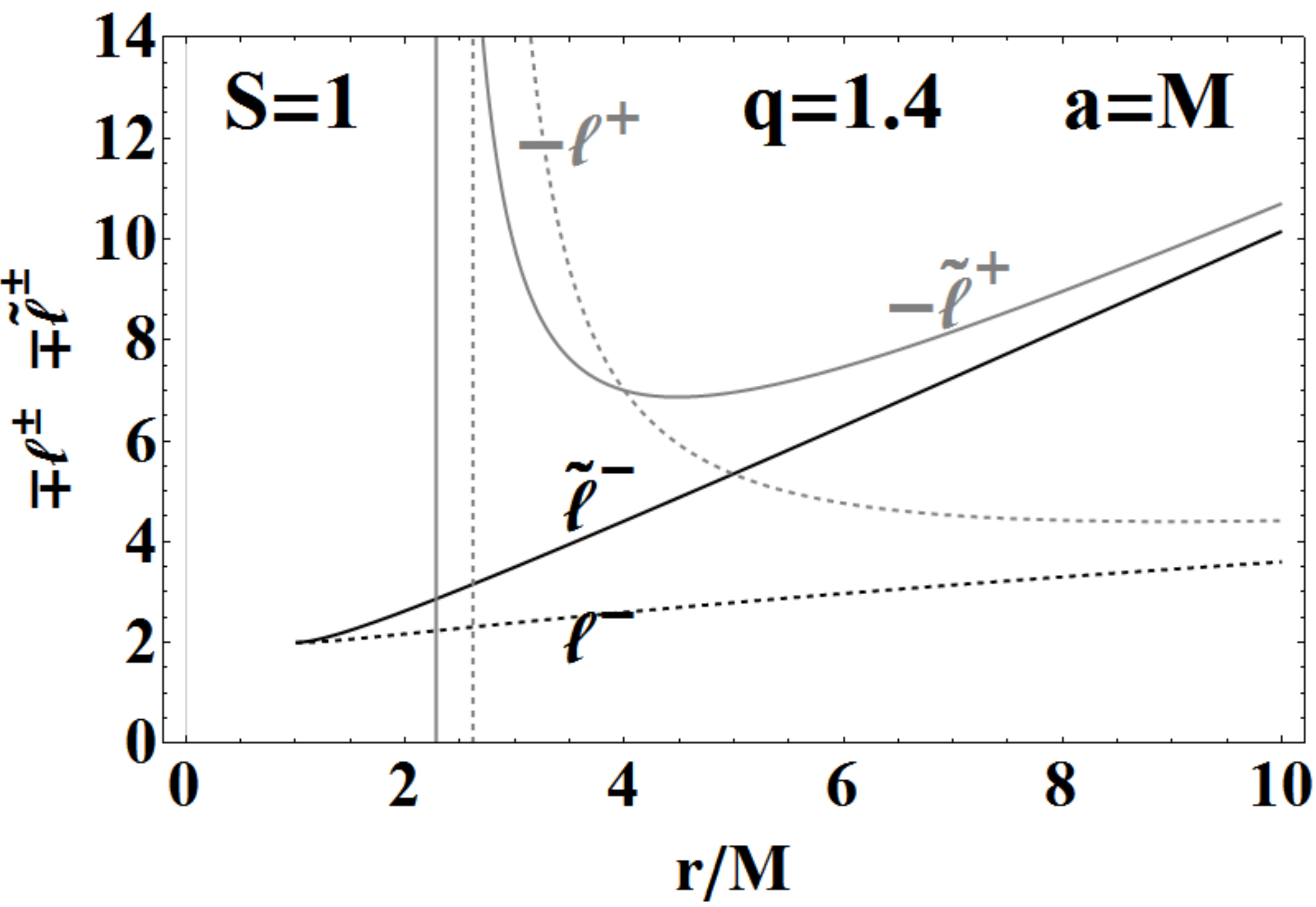}
\caption{Fluid critical angular momenta $\widetilde{\ell}_{\pm}$,  in equation\il\ref{Eq:dilde-f} for magnetized fluid. Limits $\ell_{\pm}$ for non-magnetized fluids are also shown, for fixed Kerr \textbf{BH} spin $a\in[0,M]$, magnetic parameters $\Sa$  and $q$, as function of $r/M$. The limit of the static Schwarzschild solution for $a=0$ (where $\ell_{\pm}=\ell$ and $\widetilde{\ell}_{\pm}=\widetilde{\ell}$) is also shown.}
\label{Fig:colin}
\end{figure}
\begin{figure}
%\includegraphics[width=0.4\hsize,clip]{Workother}
%\includegraphics[width=0.4\hsize,clip]{SaidSynergy}
%\\
\includegraphics[width=.81\columnwidth,clip]{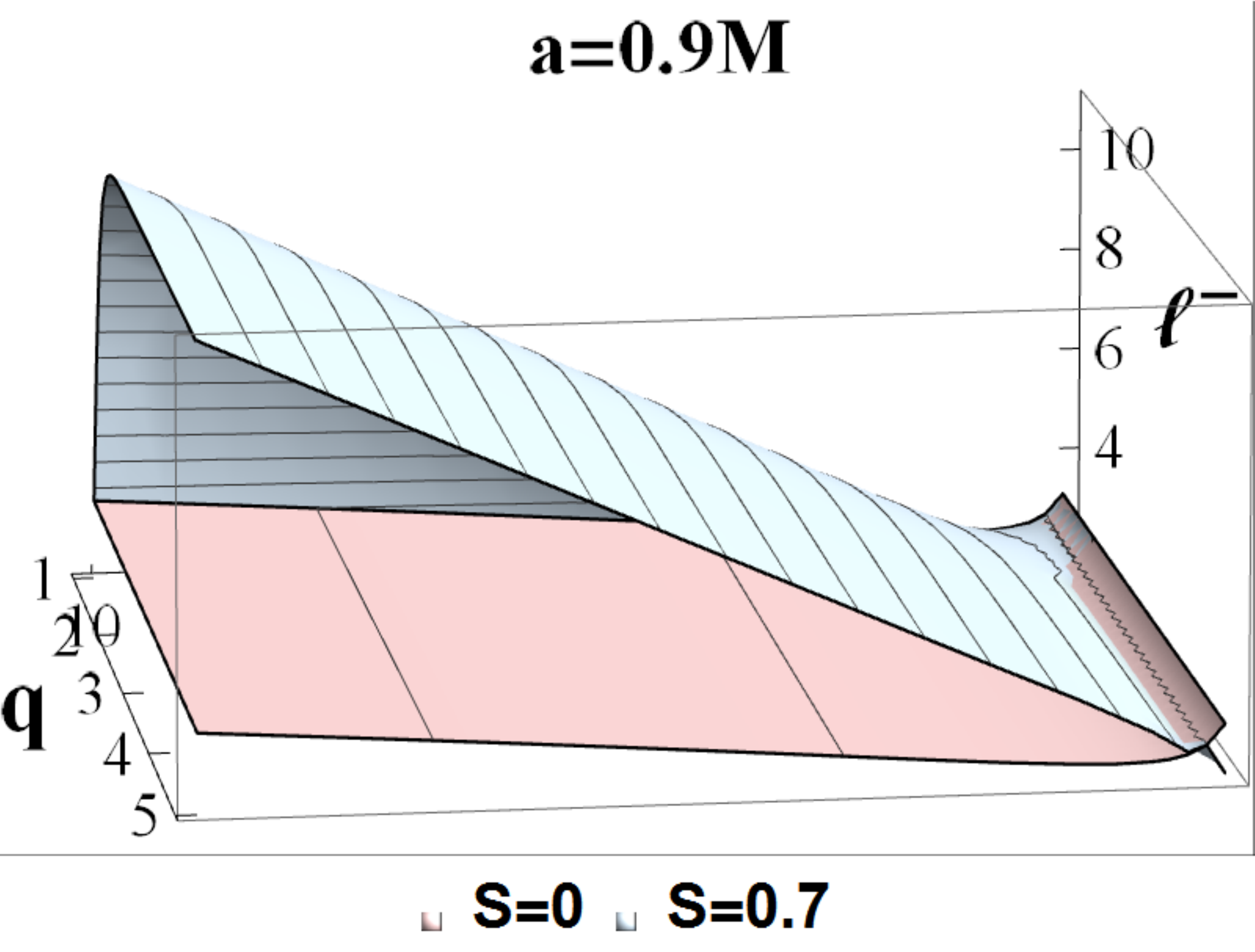}
\\
\includegraphics[width=.81\columnwidth,clip]{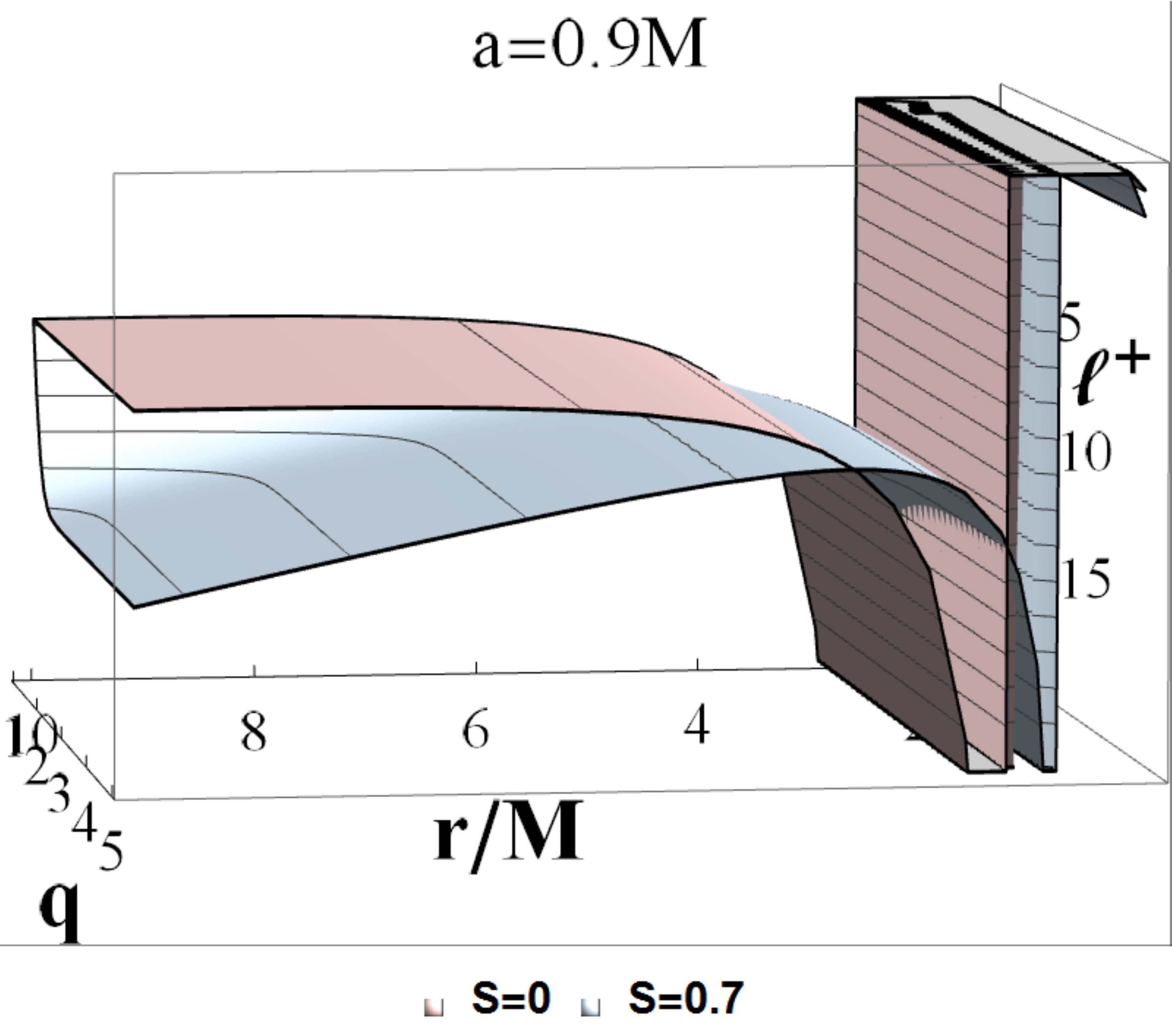}
\caption[font={footnotesize,it}]{Fluid specific angular momenta $\widetilde{\ell}_{-}$ (upper panel)  and $\widetilde{\ell}_{+}$ (below panel) for corotating and counterrotating  fluids  as function of $r/M$ and magnetic  parameter  $q$. Different values of parameters $\Sa$ are considered.}
\label{Fig:SaidSynergymat}
\end{figure}
An important part of our comparative analysis of  magnetized and non-magnetized fluids  is  grounded  on the task  to verify the presence of any \textbf{RAD} systems  constraints   induced by the magnetic field influence, whose presence here  is controlled  by  $\Sa$ and $q$ parameters, and viceversa to constraint the couple  ($\Sa$,  $q$) according to the \textbf{RAD} characterization.
It was therefore necessary to introduce an adapted function $\mathcal{\Sie}_{crit}(r;\ell,q)$,
whose level surfaces,  $\mathcal{\Sie}_{crit}(r;\ell,q)=$constant,
 provide  the values of the parameter  $\Sa$, for one or two tori.
From the  equation for the  critical points of the hydrostatic pressure, we find $\mathcal{\Sie}_{crit}(r;\ell,q)$ as follows:
{{
\bea\nonumber
\mathcal{\Sie}_{crit}\equiv-\frac{\Delta^{-\Qa}}{\Qa}\frac{a^2 (a-\ell)^2+2 r^2 (a-\ell) (a-2 \ell)-4 r (a-\ell)^2-\ell^2 r^3+r^4}{2  r  (r-1)\left[r (a^2-\ell^2)+2 (a-\ell)^2+r^3\right]}
\\
\label{Eq:Sie-crit}
\eea}}
(with $r\rightarrow r/M$ and $a\rightarrow a/M$)--see Figure\il\ref{Fig:3D}.
Firstly we note, as in equation\il\ref{Eq:dilde-f}, the explicit dependence on $\Qa=q-1$, and on  the quantities $\ell\pm a$--see also discussion in \citet{pugtot}.
We note that a negative solution for $\Sa_{crit}$ may appear  also for $q>1$ (see  Figure\il\ref{Fig:3D}, however we shall briefly consider also tori with   $q<1$  in section\il\ref{Sec:q-less-1}.
More precisely this function of the radius $r$, the parameter $q$ and  the momentum parameter $\ell$,  represents the values of $\Sa$ as a function of $r$, for  which    critical points of the function $\widetilde{V}_{eff}$ exist. In other words it provides indications on the existence of the  solutions of the Euler equation, according to our constraints, fixing  the radii  $r_{\odot}$ and, eventually, for unstable phase, the location of $r_{\times}$. The existence of  a maximum pressure point $r_{\odot}$ is sufficient to establish  whether  a  toroidal solution is possible, while   $r_{\times}$ envisages the possible deviation   of the   equilibrium condition  from  the non-magnetized case,  here the surface $\Sa_{crit}=0$. Thus, by analyzing the surfaces $\Sa_{crit}=$constant  we are able to assess in quantitative manner the magnetic field influence in the \textbf{RADs} formation and instability.

We  summarize our findings  as follows:
 generally, \textbf{RADs}  solutions are possible    when the magnetic parameters $ \Sa$ and $q$ are  balanced in such a way that their   combination    remains    small enough,  i.e. the greater is the $q>1$ and the  smaller  has to be   $\Sa$  and viceversa.
 This fact can be therefore seen  an indication of the possible effects of the magnetic field in the direction of suppressing the formation of equilibrium magnetized toroidal configurations. It is then interesting to note   the emergence of a  relation between the field parameter   $q$  and the magnitude   $\Sa$. However both this  constraint and the  range of variations for $q$ and $\Sa$ actually  depend on the \textbf{BH} spin-to-mass ratio and on the relative rotation of the orbiting fluids.
Remarkably  this analysis has proved also the different behavior of  $\ell$counterrotating and $\ell$corotating tori in presence of toroidal magnetic field, which is  also particularly evident in range $q<1$  considered here in the sideline of this investigation in section\il\ref{Sec:q-less-1}--
Figures\il\ref{Fig:Lyb-tody}.
Concerning the analysis for $q>1$,
as clear from Figures\il\ref{Fig:voice}, there is   $\Sa_{crit}\in [0, \Sa_{\max}]$,   that is $\Sa$ is
bounded below by the  non-magnetized case  and above by  a maximum value $\Sa_{\max}$, which however is not always present--see Figures\il\ref{Fig:voice}-\emph{second from above}. The situation depends mostly on the relative rotation of the fluids and also from the \textbf{BH} spin. It can be shown that  the maximum value $\Sa_{\max}$ depends linearly on $q$.
A systematic study of the  solutions in all the parameters space,  which would imply the combined selection of different ranges of parameters is  left here  for future investigation: however  we can  note that the presence of maximum for $\Sa_{crit}$ is related to  the presence of instabilities, consequently this study  provides also constraints  on the emergence of  P-W instability  and on the relevance of the toroidal   magnetic fields contribution in enhancing accretion.   Figures \ref{Fig:voice}
 show some exemplary cases.
Integrations of Euler equation  for a couple of magnetized tori  in fixed spacetimes are   in Figures\il\ref{Griegn}.
As mentioned above, these analysis confirm the requirement of small values of   $q\Sa$,  it follows that the magnetic field is strongly constrained by the torus formation: the presence of a strong field in the early age of accretion disk evolution would act in direction to suppress the torus formation--Figures \ref{Fig:voice}.
Comparing then with the $\Sa=0$ case, we see that the instability points shift   away from the central attractor   which implies that the magnetic field has essentially in general  a destabilizing effect on the configurations, fostering the instability emergence.
Then the accreting magnetized tori are   generally smaller (equatorial plane elongation) than the non magnetized ones.
 Asymptotically, for large values of  $r$,  the parameter  $\Sa$ decreases to zero values for $q>1$. In particular, this means that a magnetized torus may form    close to the central attractor.
Focusing on the $\ell$corotating couples, %as shows in Figures\il\ref{} and Figure\il\ref{}-left
we note that $\Sa$  increases  with  the magnitude of  $\ell$  moving the torus and the maximum of $\Sa$ inwardly. This trend  is due to  the  coupling between  the centrifugal  and the magnetic field component of the force balance in  Euler equation, encoded in the effective potential function $\widetilde{V}_{eff}$ in equation\il\ref{Eq:goood-da}.
 The  magnetic pressure in   equation\il\ref{RSC} is independent from the fluid specific angular momentum $\ell$, viceversa  the magnetic field $B_{\phi}$ explicitly depends on  $\ell$ through  $\la$ in equation\il\ref{Eq:Kerr-case}.
The greater is the fluid rotation  and the greater is the   magnetic  field, increasing the maximum $\Sa_{crit}=\Sa_{\max}$  values    and the radius $r_i:\; \Sa_{crit}(r_i)=\Sa_{\max}$  for  the case of $\Sa=0$. This behavior is   substantially independent from the   sign of rotation  with respect to the central attractor- for the corotating and counterrotating couples of $\ell$corotating tori- see Figures\il\ref{Fig:voice}. Fluid rotation would  act therefore so  to offset the effects of the magnetic  field.
Considering then the  $\ell$corotating  fluids,  \textbf{RADs} may form at  equal $q$  and $\Sa$ --constant lines in $\Sa_{crit}$ in Figures\il\ref{Fig:voice}-Figures\il\ref{Fig:3D}. This would  be  an  important indication in support of  the \textbf{RADs} origin, with constrained angular momentum, from one  common   embedding material   as envisaged in \citep{ringed,open,dsystem}. The magnetic field would act so as to foster the formation of a  single accretion disk, following  the emergence of \textbf{RAD} instability  originating from each torus unstable phases or  from tori collision.
 It is clear then that in the $\ell$corotating couples,  the maximum common value of the parameters $\Sa$ for the tori to be considered is $\Sa=\Sa_i$ relative to the inner torus.

The $\ell$counterrotating magnetized fluids constitute a particular
 interesting case where, for
  $\Sa=0$,  couples $\cc_{\times}^+<\cc^-$ and   $\cc_{\times}^-<\cc_{\times}^+$, might occur.
As clear from Figures\il\ref{Fig:voice} the following two cases may occur:
\textsl{\textbf{i.}}  there is partial or total overlapping of the curves $\Sa_-$ (for the inner torus) and $\Sa_+$, and in this case the situation for a \textbf{RAD} is analogue to the $\ell$corotating case discussed above.
\textsl{\textbf{ii.}}  The second case consists of  $\Sa_{crit}$  curve profiles which are totally disjoint, as in  Figures\il\ref{Fig:voice}--\emph{third line}.
In general  in the  $\ell$counterrotating case  we can assess the  different coupling between the centrifugal component and the magnetic field contribution  in the counterrotating and corotating cases respectively (for  $a\neq0$) (and this is especially clear for the case  $q<1$, which is also addressed in  Sec\il\ref{Sec:q-less-1}).
 The presence of a toroidal  magnetic field would distinguish between corotating and counterrotating fluids,
favoring  the formation of the first  \citep{Volonteri:2002vz}.
In the  $\ell$counterrotating couples, if there is no maximum $\Sa_{\max}$  the curves are overlapped as they are always in the $\ell$corotating case,  implying  that tori at equal $\Sa$ and $q$ are always possible, and this  may support the possibility of common origin for the tori.
In the case of disjointed curves,  the common $\Sa_{crit}$ parameter would be determined by the external    counterrotating torus. This suggests  a different  origin for the  tori of an $\ell$counterrotating couple in the case $\cc_-<\cc_+$ only \citep{dsystem}.
\begin{figure}
\centering
\includegraphics[width=0.91\hsize,clip]{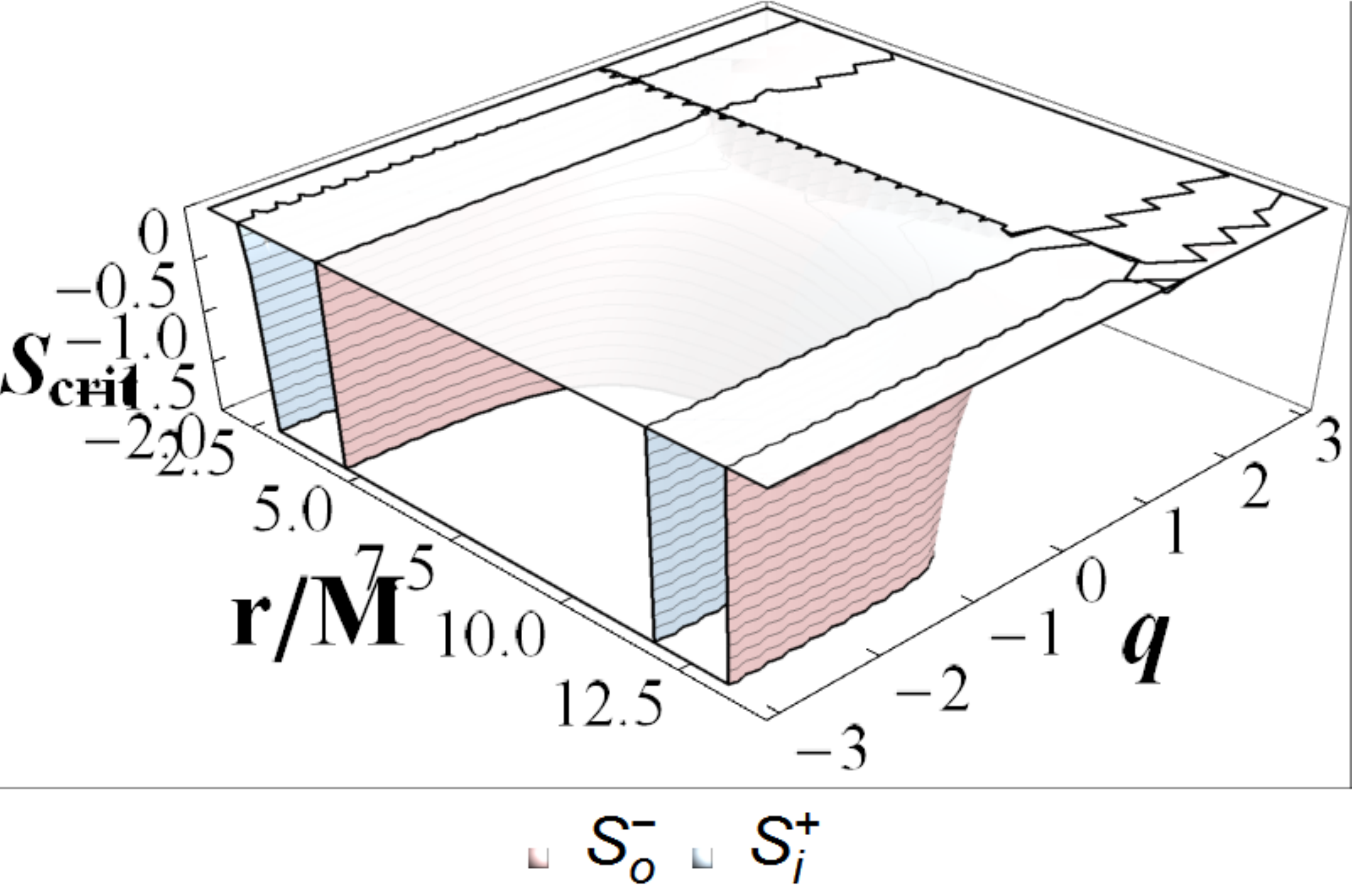}
\\
\includegraphics[width=0.91\hsize,clip]{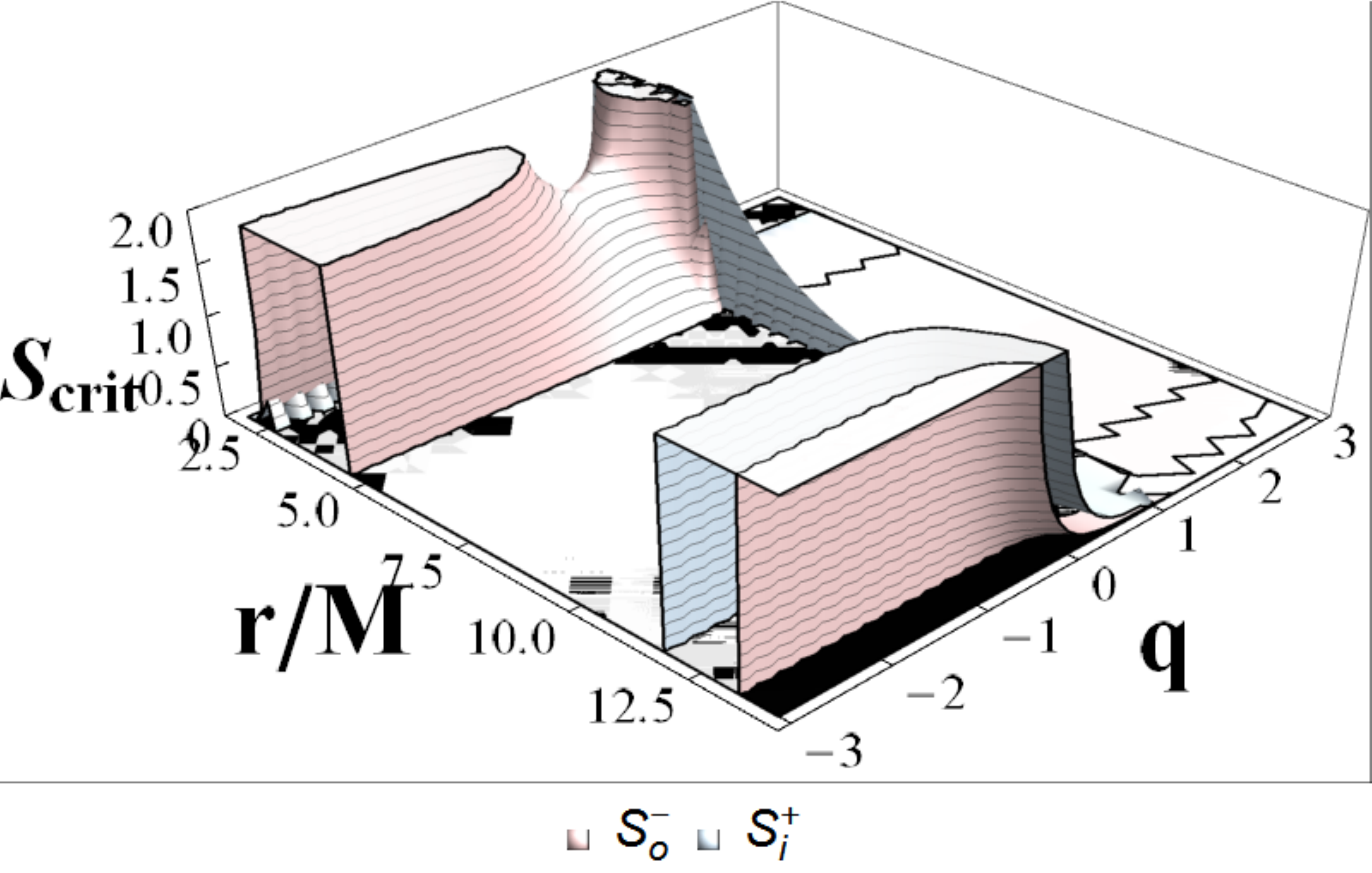}
\caption[font={footnotesize,it}]{Function $\Sa_{crit}(r; a, \ell, q)$ introduced in equation\il\ref{Eq:Sie-crit} as function of radius $r/M$ and magnetic parameter $q$, for fluid  specific angular momentum ${\ell}_{\pm}$ (equation\il\ref{Eq:dilde-f}) for $\Sa_{\pm}$ respectively according to fluid rotation. The \textbf{BH} spin is $a=$, fluid specific angular momentum $\ell^-=4.01$ and $\ell^+=-4.4$, $a=0.382M$, for negative (positive)  values of  $\Sa_{crit}(r; a, \ell, q)$ upper panel  (below panel). See also Figures\il\ref{Fig:voice}).}
\label{Fig:3D}
\end{figure}
\begin{figure*}
\includegraphics[height=4cm, width=0.4\hsize,clip]{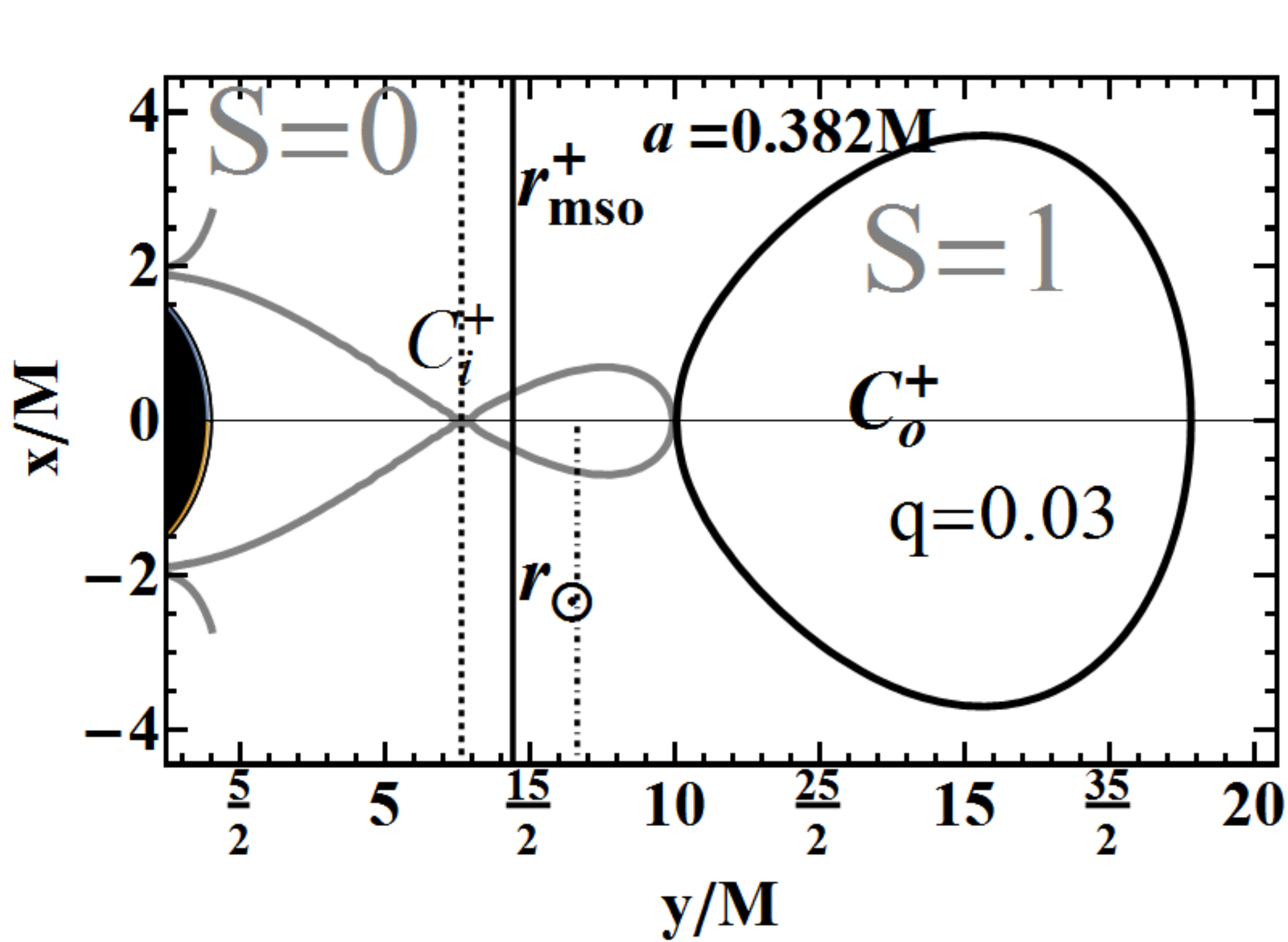}
\includegraphics[height=4cm, width=0.4\hsize,clip]{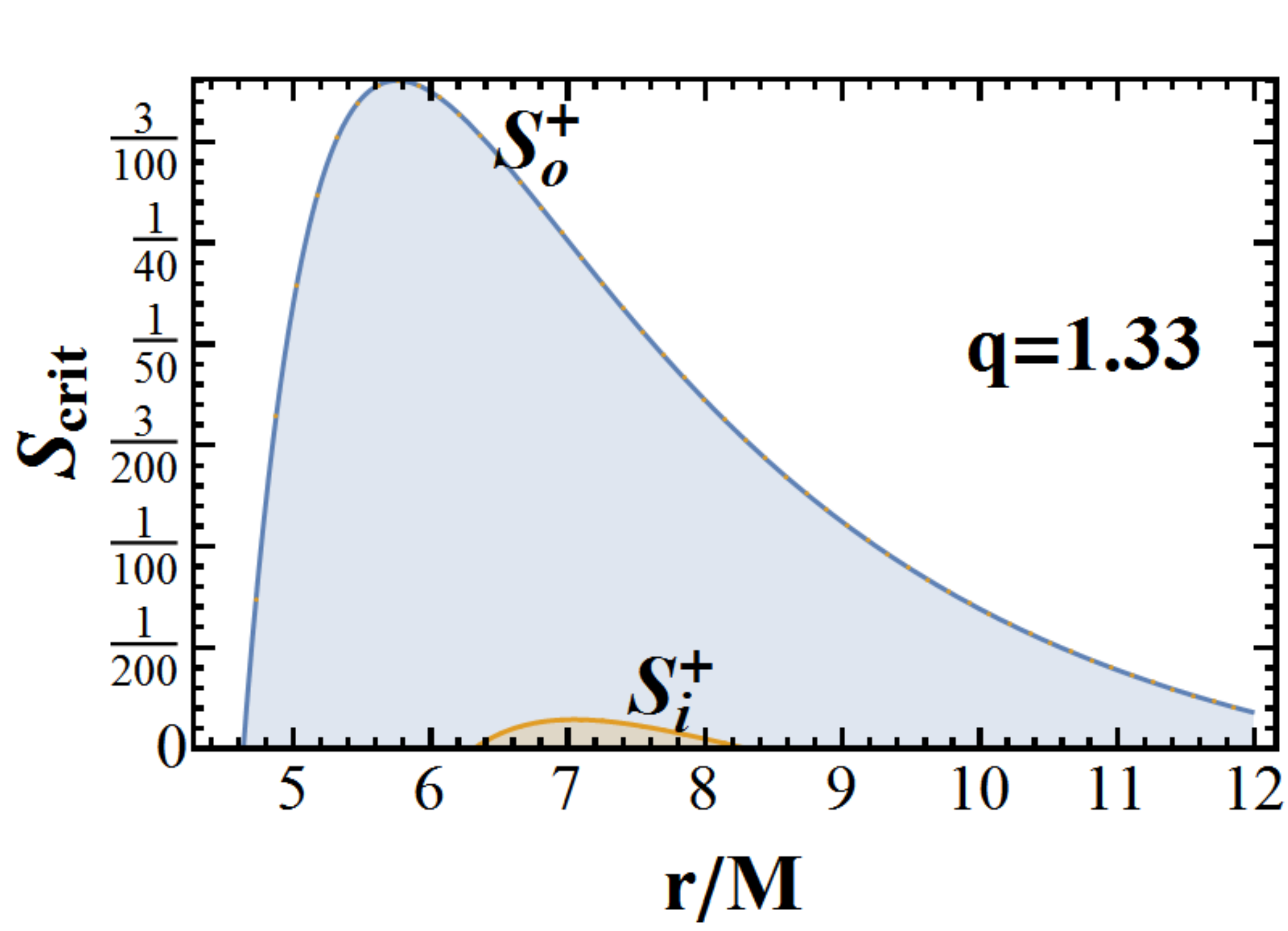}
\includegraphics[height=4cm, width=0.4\hsize,clip]{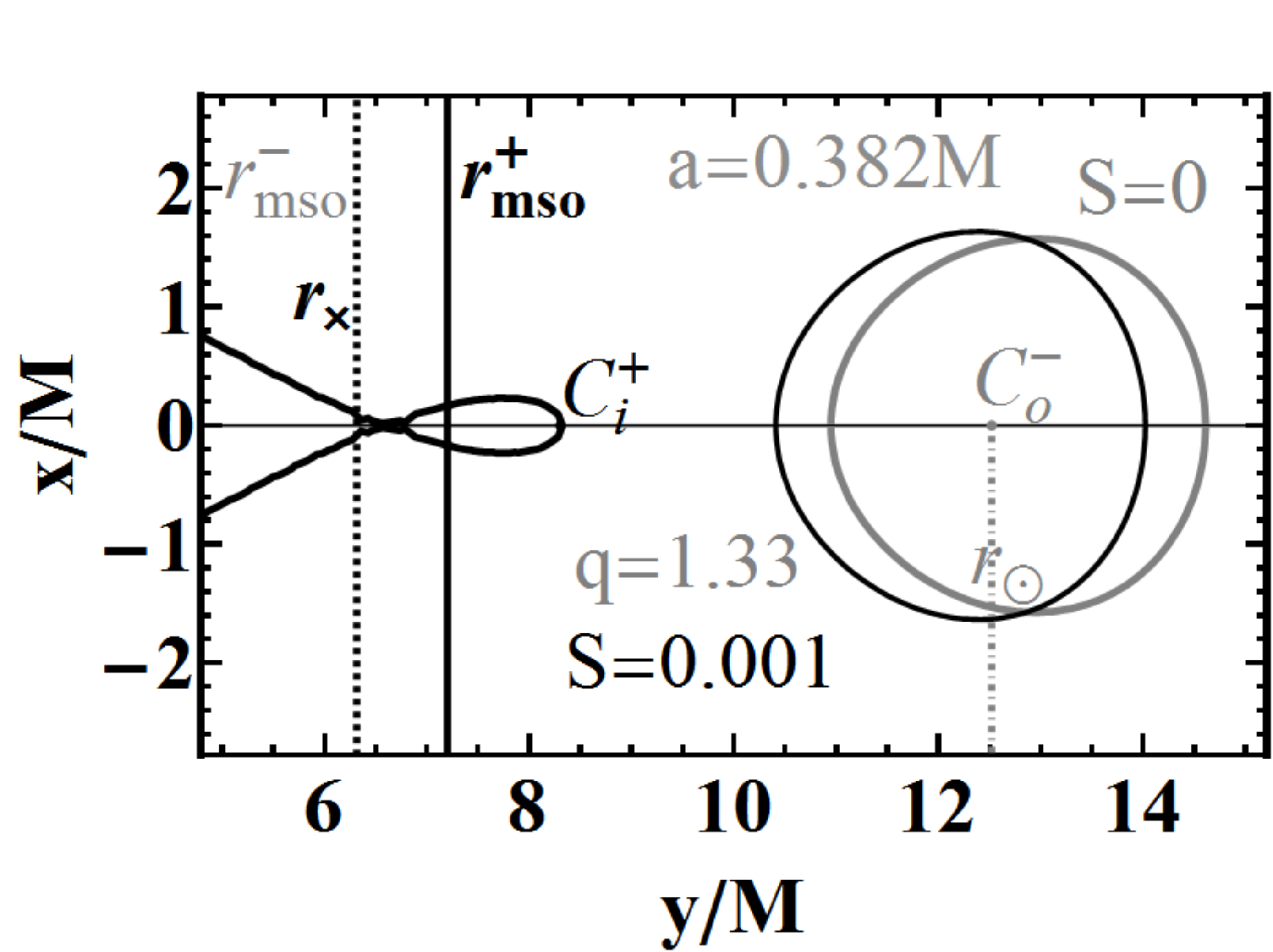}
\includegraphics[height=4cm, width=0.4\hsize,clip]{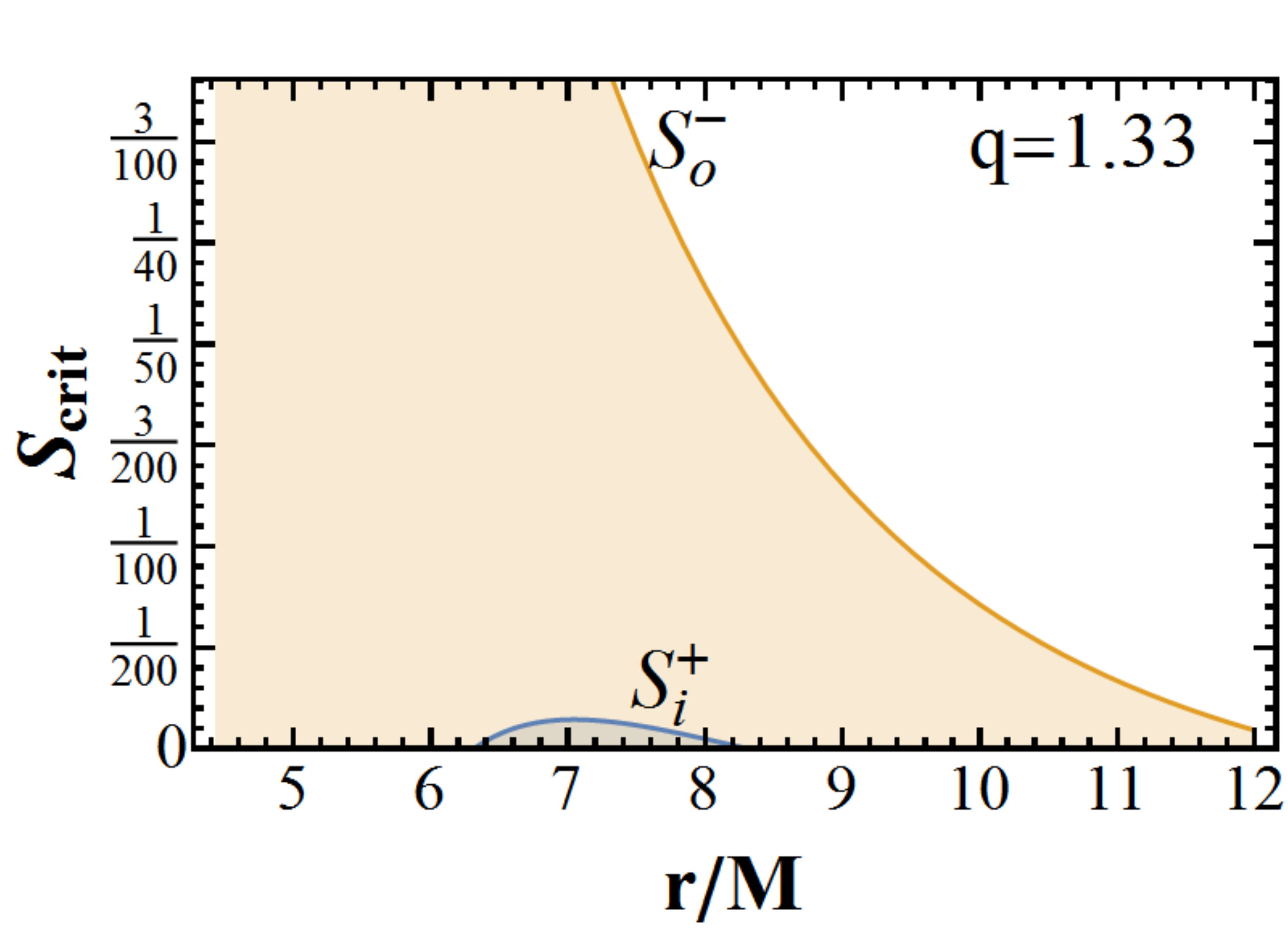}
\\
\includegraphics[height=4cm, width=0.4\hsize,clip]{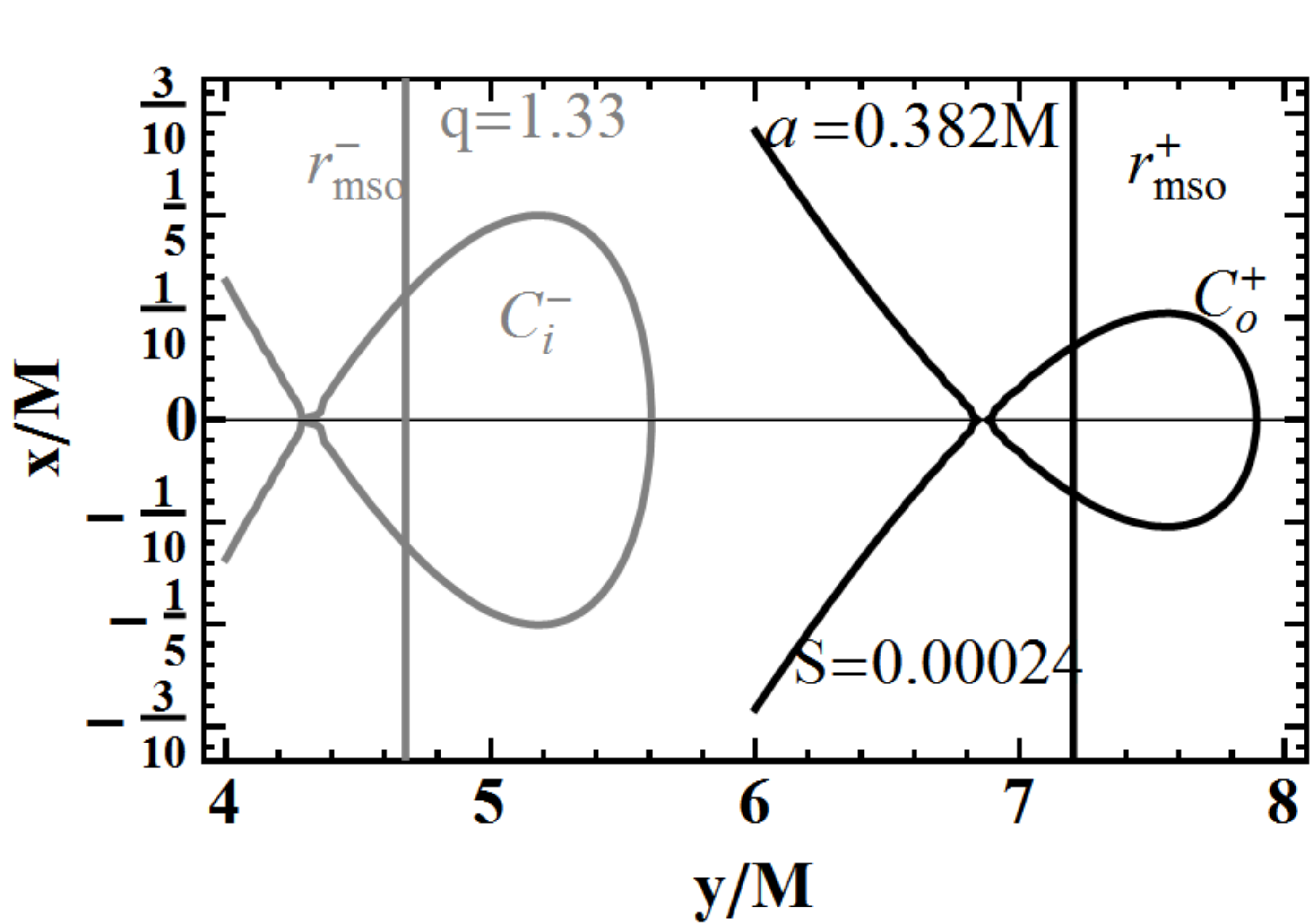}
\includegraphics[height=4cm, width=0.4\hsize,clip]{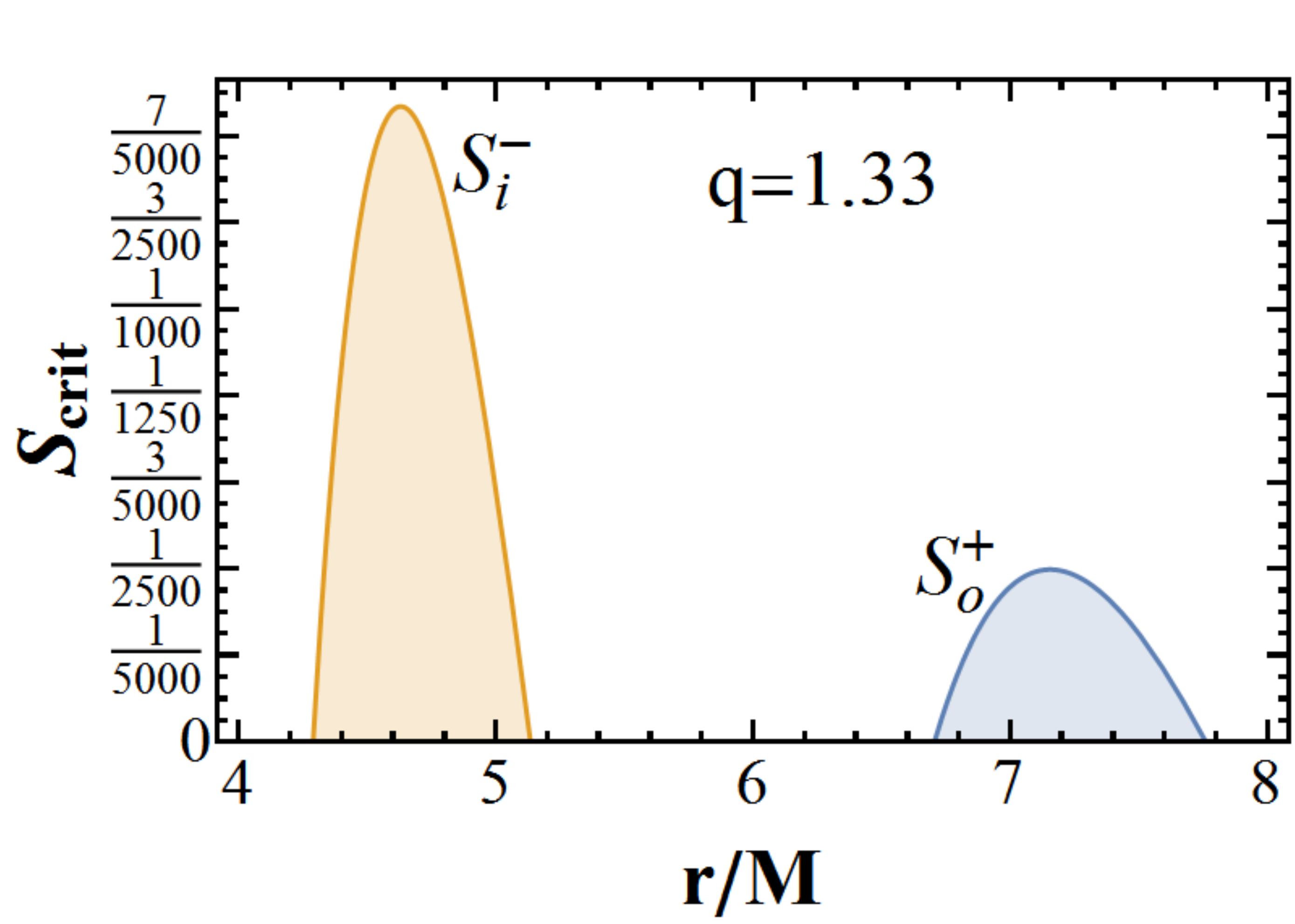}
\includegraphics[height=4cm, width=0.4\hsize,clip]{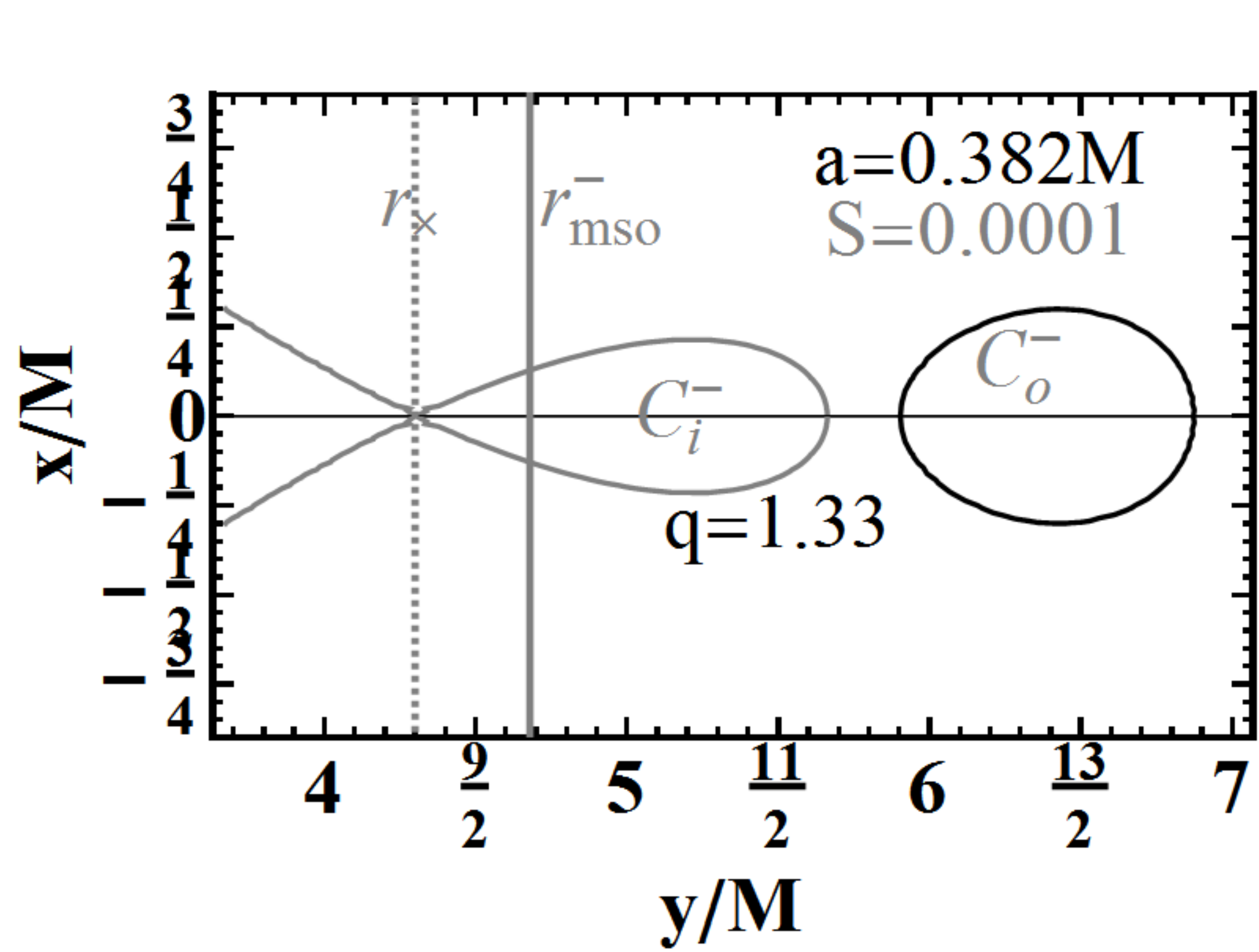}
\includegraphics[height=4cm, width=0.4\hsize,clip]{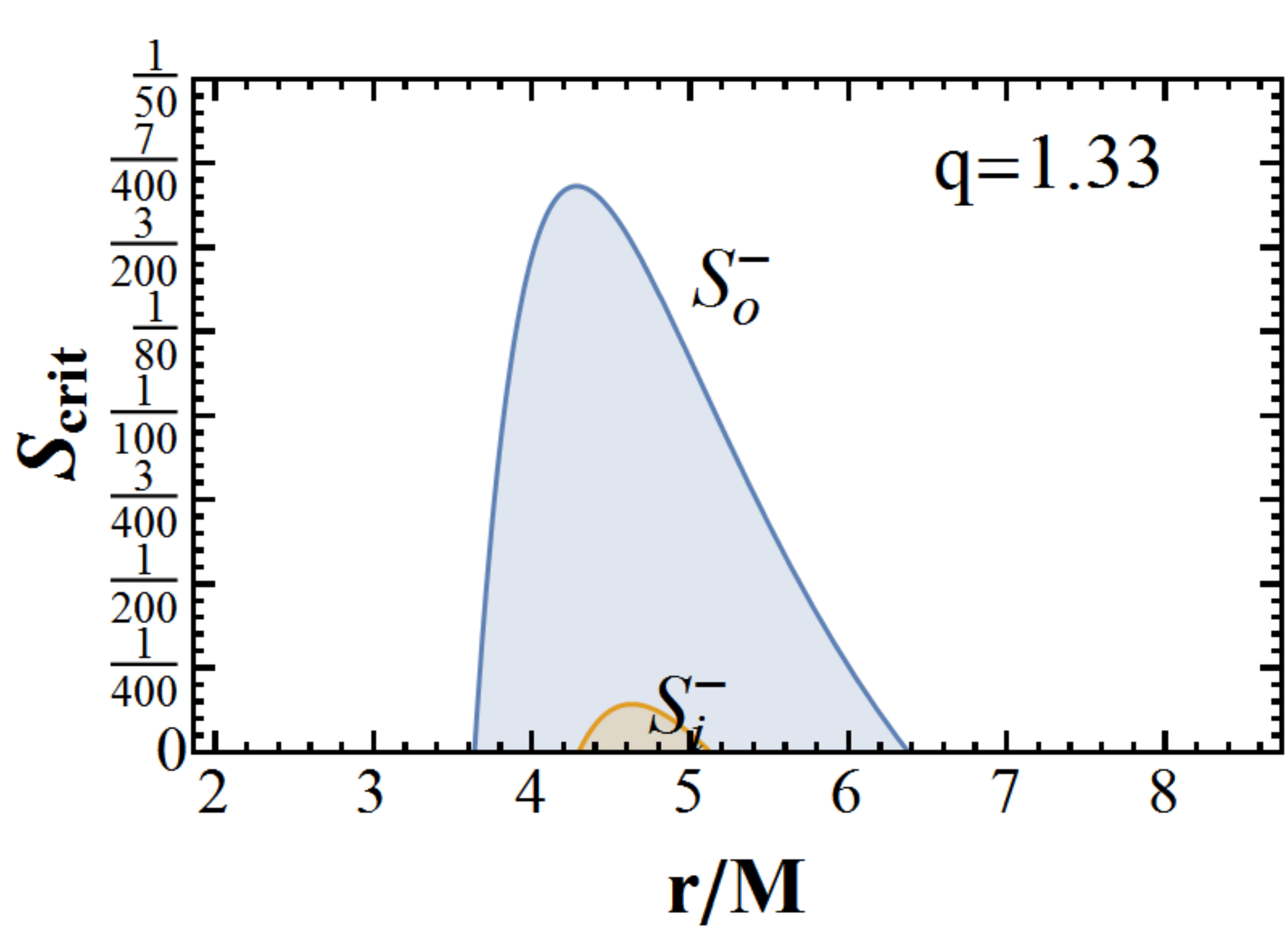}
\\
\includegraphics[height=4cm, width=0.4\hsize,clip]{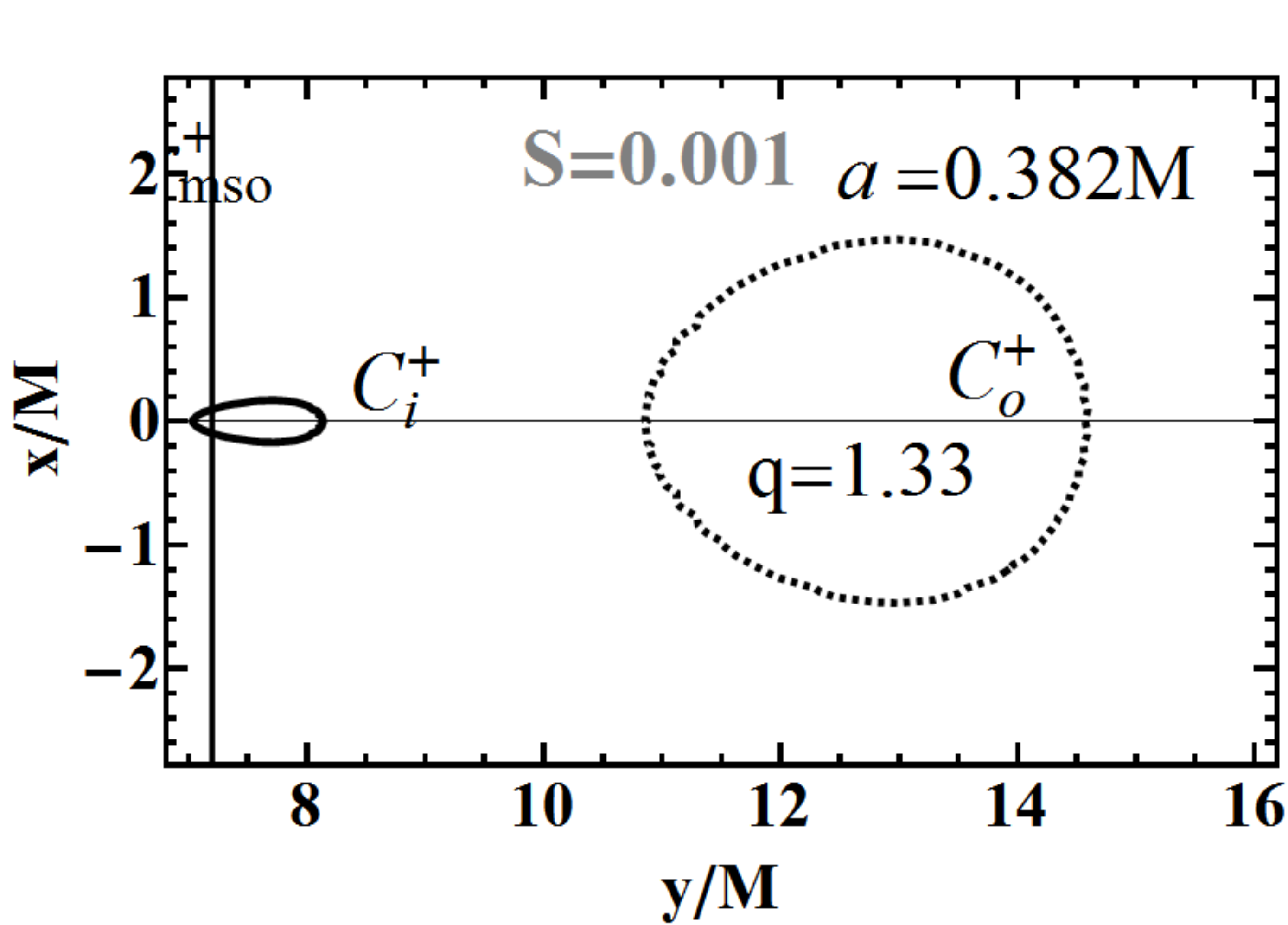}
\includegraphics[height=5cm, width=0.4\hsize,clip]{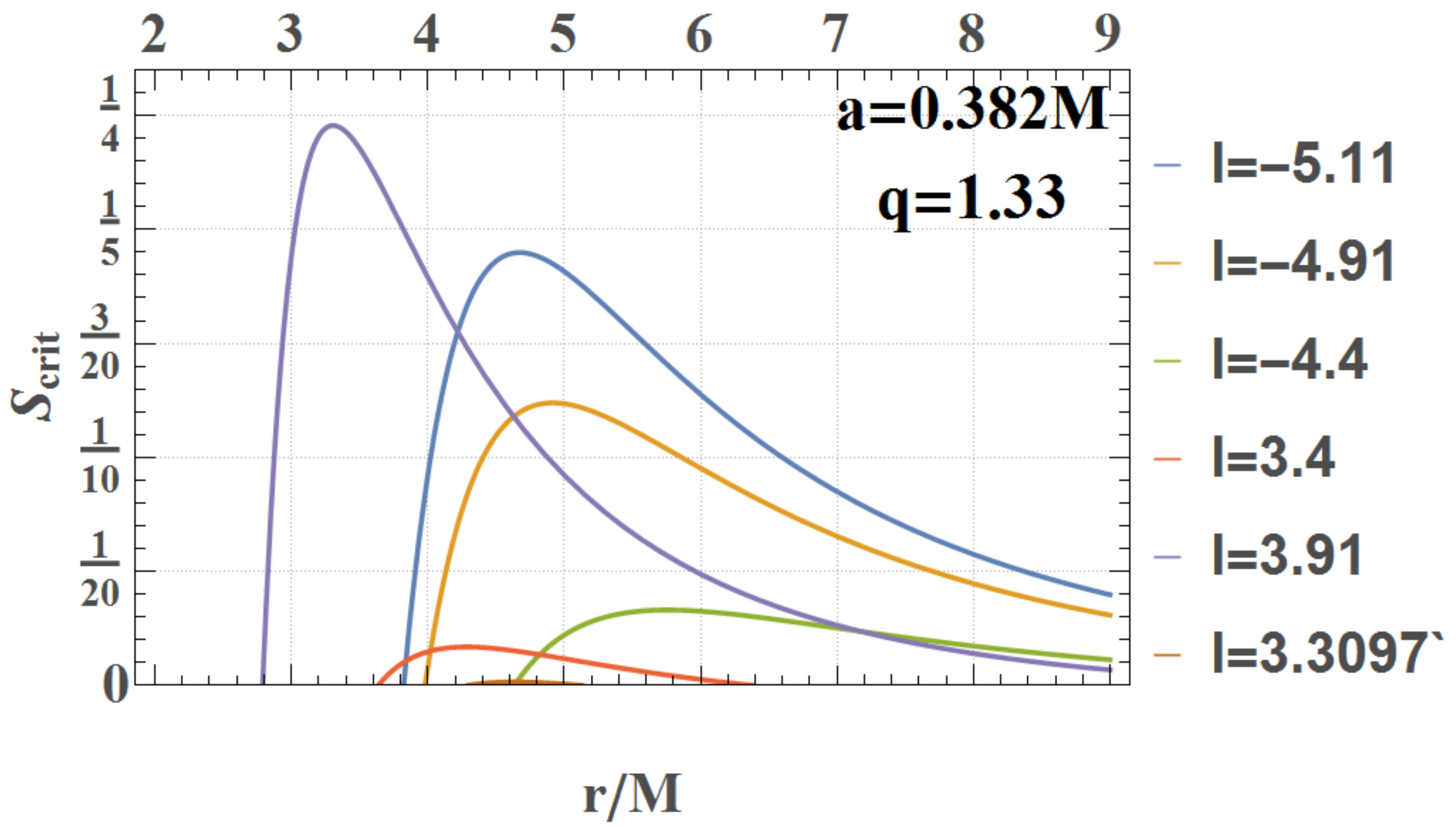}
\caption{Left: Cross sections, on the equatorial plane, of the outer Boyer surfaces   (Roche lobes)   for  $\ell$counterrotating and $\ell$corotating  tori  orbiting a central Kerr \textbf{BH} (left and bottom panel),
 and associated $\Sa_{crit}$  parameter (right panels) as function of $r/M$--see also equation\il\ref{Eq:Sie-crit}. $(x, y)$ are Cartesian coordinates.  Bottom panel: $\Sa_{crit}(r; a, \ell, q)>0$ for different values of momentum $\ell_{\pm}$.}
\label{Fig:voice}
\end{figure*}
\subsection{Some  considerations on the parameter choice}\label{Sec:q-less-1}
In this section we further discuss  the parameter choice focusing on the  range of variation for the $q$-parameter.
We have seen from equations\il\ref{Eq:Kerr-case}-\ref{Eq:goood-da} and equations\il\ref{Eq:dilde-f}-\ref{Eq:Sie-crit}, that the  \textbf{RADs}   strongly depend on the   parameter $\mathcal{Q}=q-1$. We assumed $\mathcal{Q}>0$, with  $\mathcal{Q}=0$ matching  the limiting case for null magnetic component.
 Considering again Euler equation \ref{Eq:scond-d}, with the  effective potential $\tilde{V}_{eff}$ in equation\il\ref{Eq:goood-da} including the magnetic contribution, we can note that  $\mathcal{Q}$ parameter in fact    defines positive or negative contribution of the magnetic pressure  in the pressure-force balance, together with the barotropic pressure contribution and the centrifugal  and gravitational parts, included in the (non-deformed) effective potential $V_{eff}$.
In here we briefly discuss  results of the  analysis performed in this extended parameter ranges considering a negative $\mathcal{Q}$--see Figures\il\ref{Fig:Lyb-tody}.

We can clearly see the presence of maxima and of possible negative values (for $q <1$) of the  parameter\footnote{This special choice of $\Sa$ ($\mathcal{M}$) and $q$ parameters  requires a throughout discussion of the   matter and fields characteristic as described by these values, and the implication on the conservation equations, the  Komissarov field  and, importantly, the \textbf{RADs} components boundary conditions, which are here particularly relevant as excretion disks may appear. This analysis is left for future investigation. However, without over-deepening this aspect that eludes the purposes of the present  analysis, we can say that a study of the $\Sa_{crit}$ quantity  as in Figures\il\ref{Fig:3D} reveals a far more rich scenario  then the cases  depicted in Figures \ref{Fig:voice}  and Figures\il\ref{Fig:Lyb-tody}; we can consider the negative $\Sa_{crit}$ values, defined in  equation\il\ref{Eq:Sie-crit},  giving rise to accretion disks or, negative $\mathcal{Q}$, which gives rise to toroidal solutions of Figures\il\ref{Fig:Lyb-tody}  satisfying the   requirement of constant (magnetized) potential.} $\mathcal{Q}$-- Figure\il\ref{Fig:Lyb-tody}
The first  relevant  feature in this new set-up consists in the possible formation of a  multi-tori  where both  \textbf{RAD} accretion disks  have the same    $\ell$ and $K$ values, which obviously cannot be  in the case of different magnetic parameter values. This implies that each torus is not uniquely defined, in general, in  this range values  for  the magnetic parameter, by the only  fluids rotation $\ell$ and density $K$ parameters. Moreover, this suggests also that the same original matter, constituting the primordial    embedding of the disks,   may probably give rise to two different  accretion tori with equal centrifugal ($\ell$), density $(K)$  and  magnetic $(\mathcal{Q})$ properties, eventually  pointing  out   an  interesting  mechanism in the disk formation.
 The second,  relevant difference with respect to the other \textbf{RADs} considered   here,  consists in the fact that, for  the inner torus, so-called \emph{\emph{excretion}} phase  is  possible.
This mechanism  of the accretion disks instability is  indeed a well known feature of different scenario as in  \citep{Stuchlik:2014jua,[68],PPT,[70],Hana,PSzS,adamek}.
In excretion disks the balance of forces is such that the flow starts from the center of the disk and exits the outer margin (in this sense we could say there is  a role shift  between the outer  $r_{out}$ and the inner $r_{in}$ edge). Excretion disks form for example at stars merging.
However, in  all the   different circumstances considered in the former studies, a repulsive effect in the force balance appears, generally inherited as a  peculiar feature of the background  geometry, therefore enucleated in the gravitational part contribution  of the effective potential in the forces-balance equation. In our case, in this extended range of the $\mathcal{Q}$ parameter, interestingly,  the  repulsive effect is in fact introduced directly in the balance of the forces, due to the magnetic  field contribution.
These multiple configurations, seen as a very special  subgroup of \textbf{RADs}, would emerge  for the  counterrotating configurations only. This situation  leads us to  conjecture that a general classification of balance of forces in tori  may be done,
answering to the question of how should be the  effective potential modified to envisage such kind of special multi-tori with  equal parameter values  and where excretion processes may occur.
Then, the existence of these solutions,  will let us to conclude that existence of any excreting  tori in fact  may not  be exclusively attributed  to the effects of a \emph{geometric repulsive force}, due to a cosmological constant contribution \citep{[68],PPT,PSzS,[70]}, to  some kind of quantum distortion effects having  such impact on the larger  scales \citep{Stuchlik:2014jua}, or the presence of super-spinning sources \citep{adamek,Stuchlik:2013yca,Stuchlik:2010zz,Stuchlik:2011zza} for example.
These orbiting tori may  therefore represent more common situations in Astrophysics, then following  assumptions on   very special and exotic backgrounds. Particularly they may be  relevant in the early phases (as transient stages) of the accretion disks formation.
This hypothesis encourages  for future analysis  directed towards the investigations of these cases.
\begin{figure*}
\centering
\includegraphics[height=4cm, width=0.4\hsize,clip]{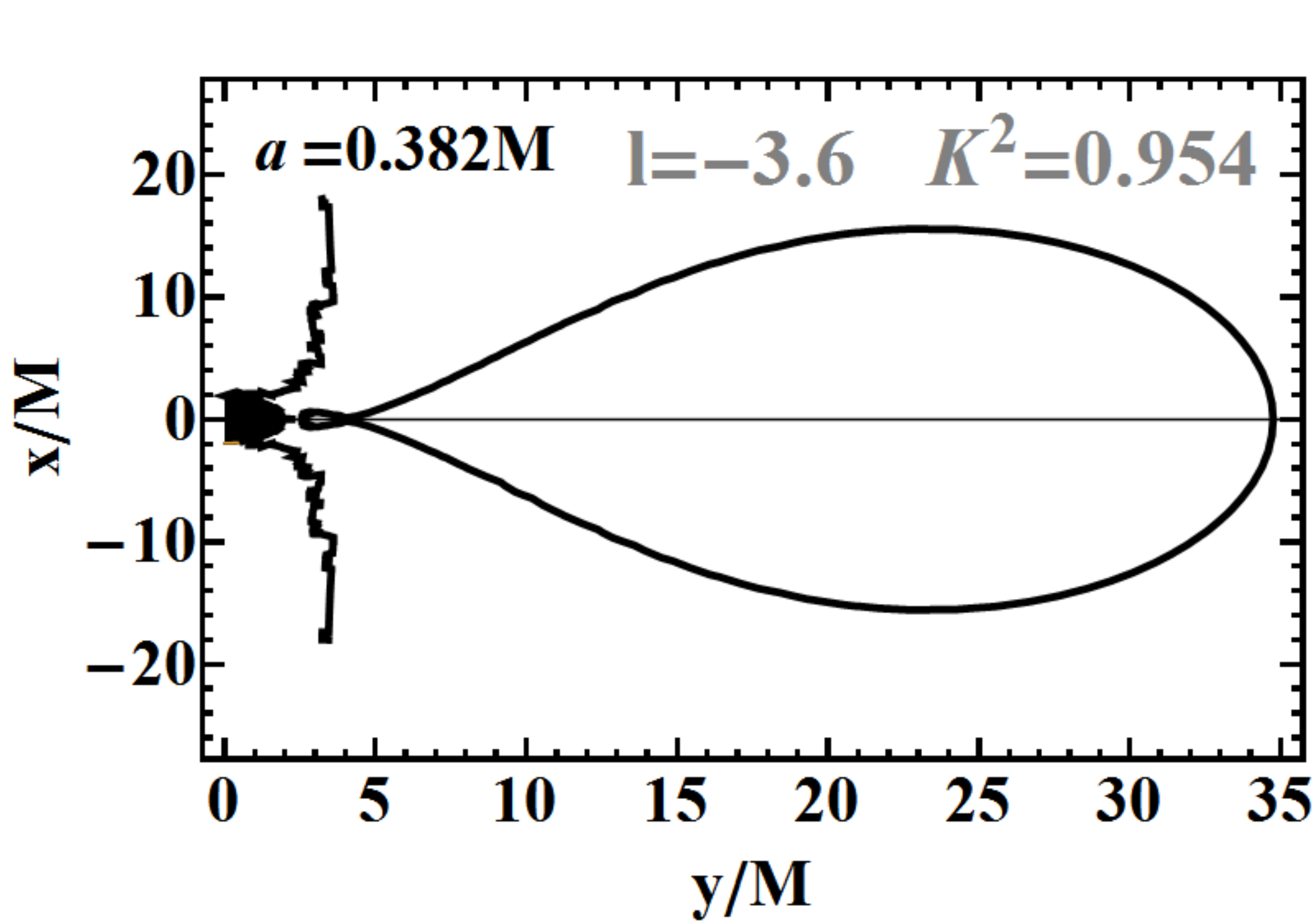}
\includegraphics[height=4cm, width=0.4\hsize,clip]{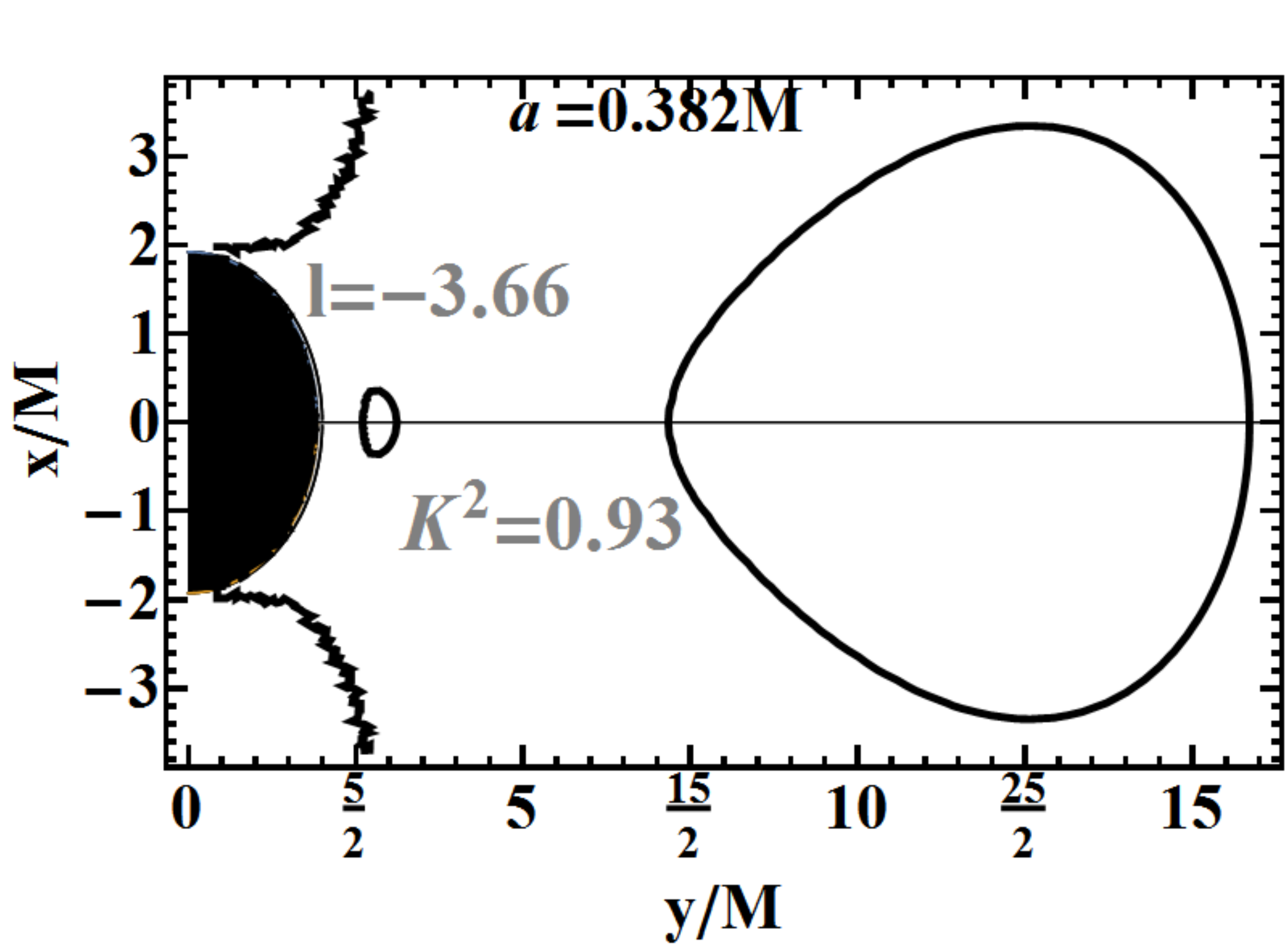}
\includegraphics[height=4cm, width=0.4\hsize,clip]{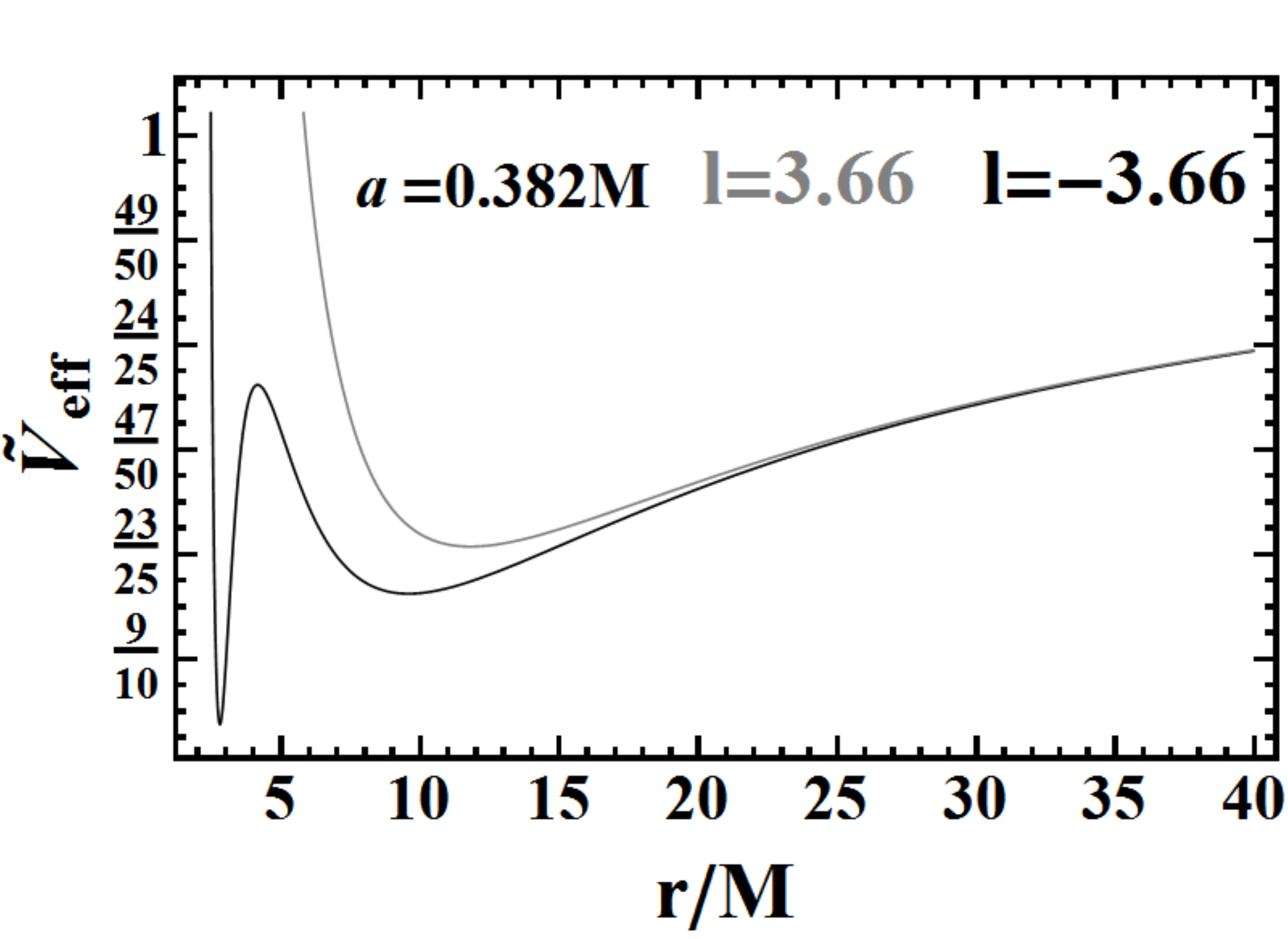}
\includegraphics[height=4cm, width=0.4\hsize,clip]{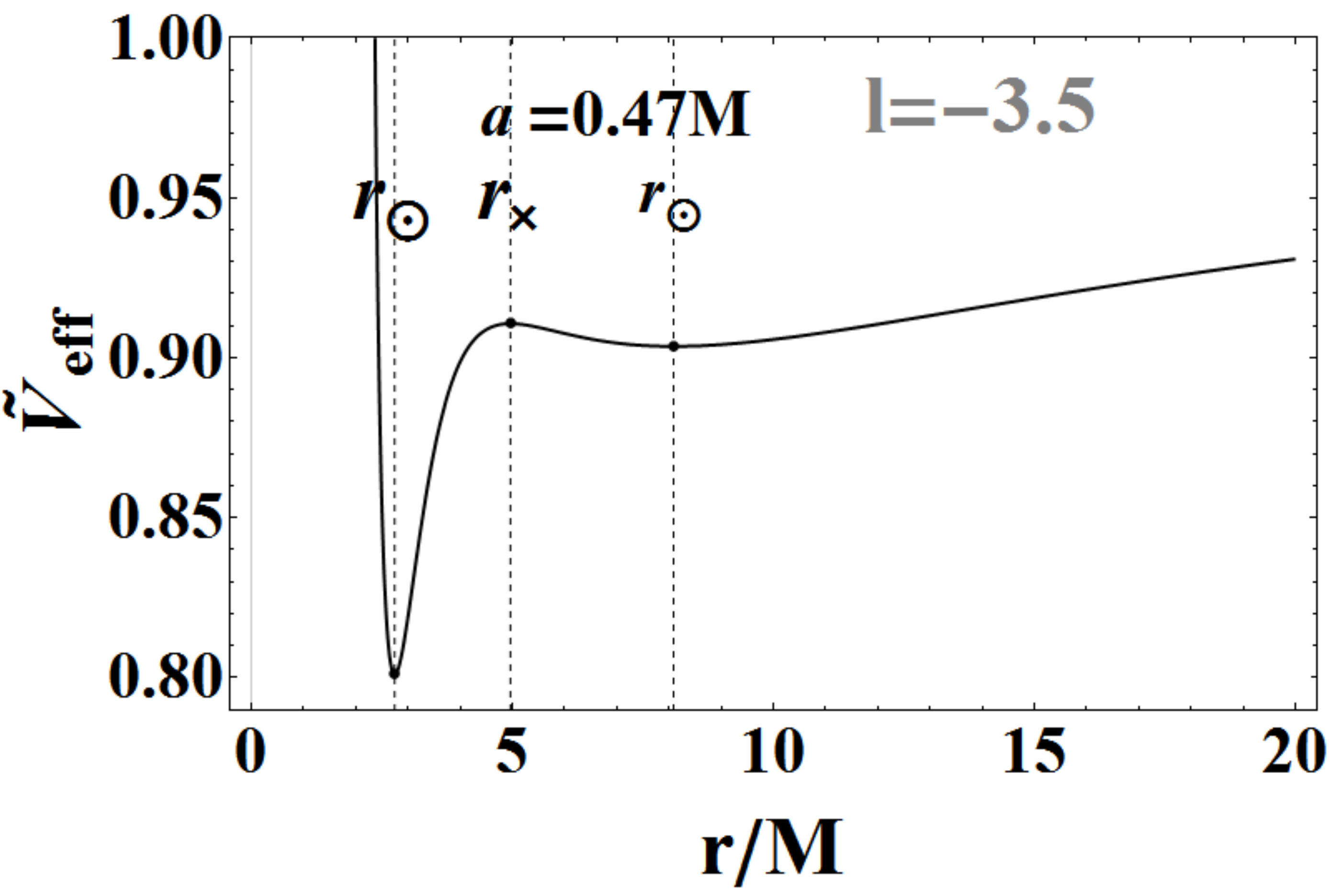}
\\
\includegraphics[height=4cm, width=0.4\hsize,clip]{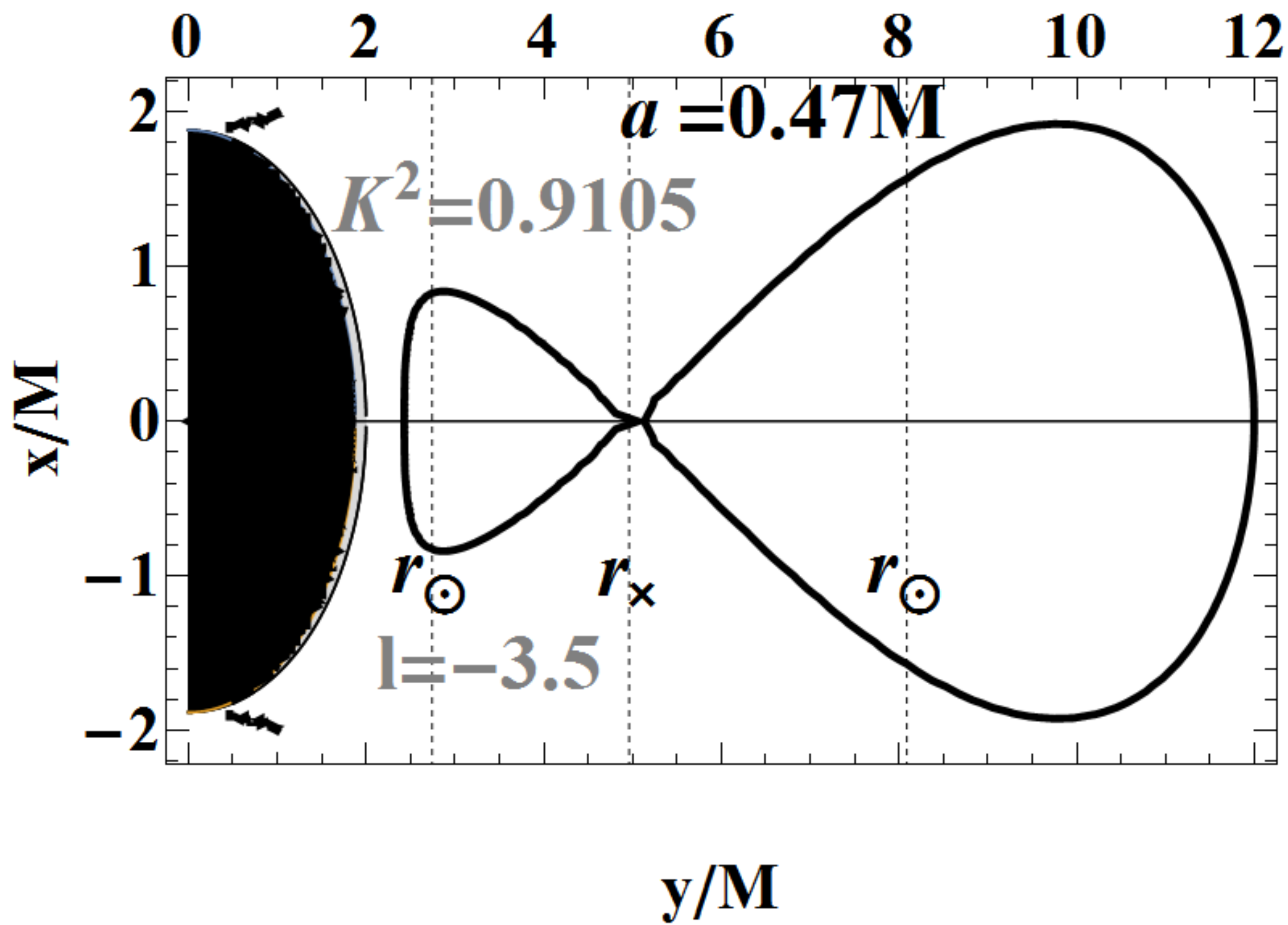}
\includegraphics[height=5cm, width=0.5\hsize,clip]{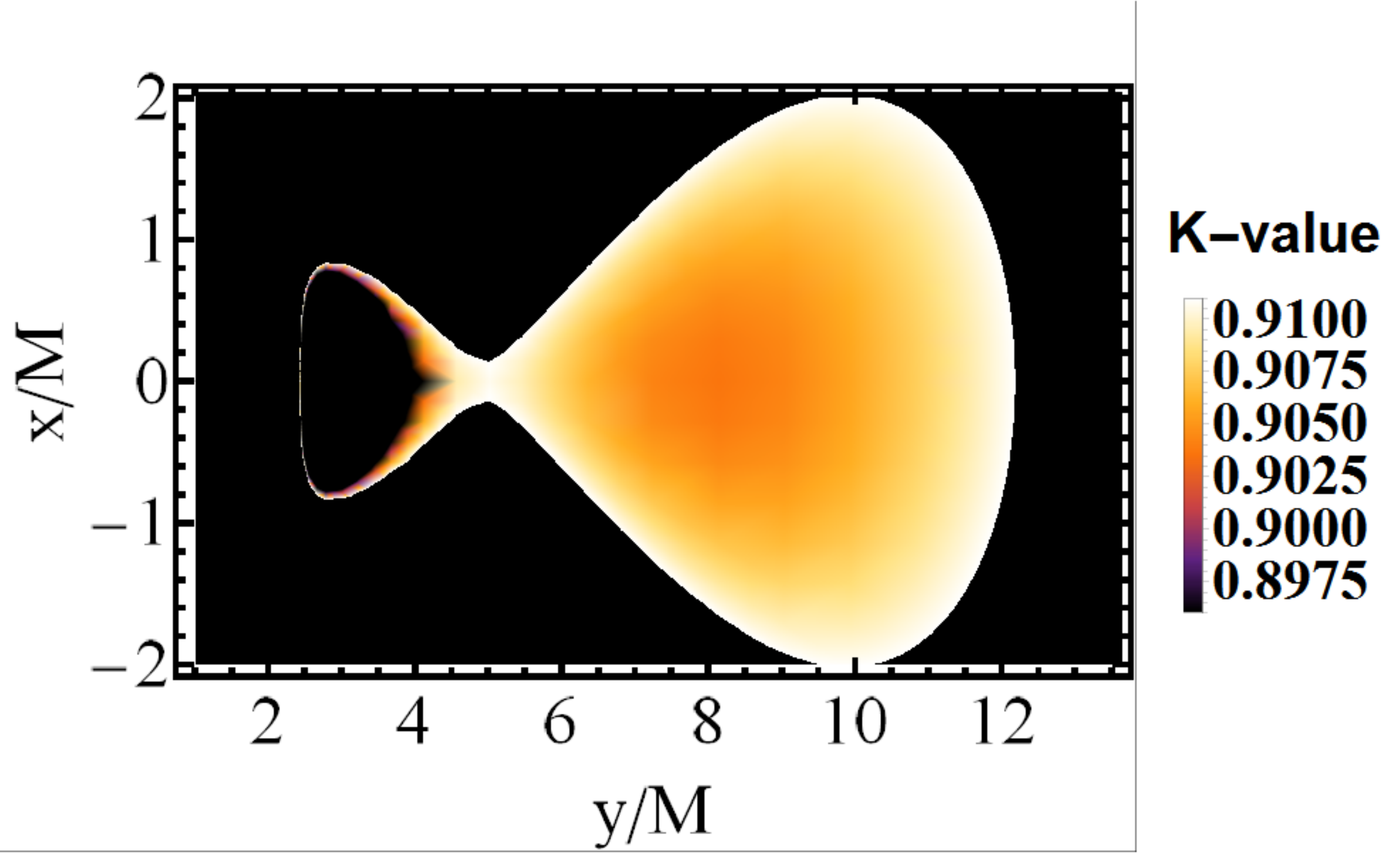}
\includegraphics[height=4cm, width=0.4\hsize,clip]{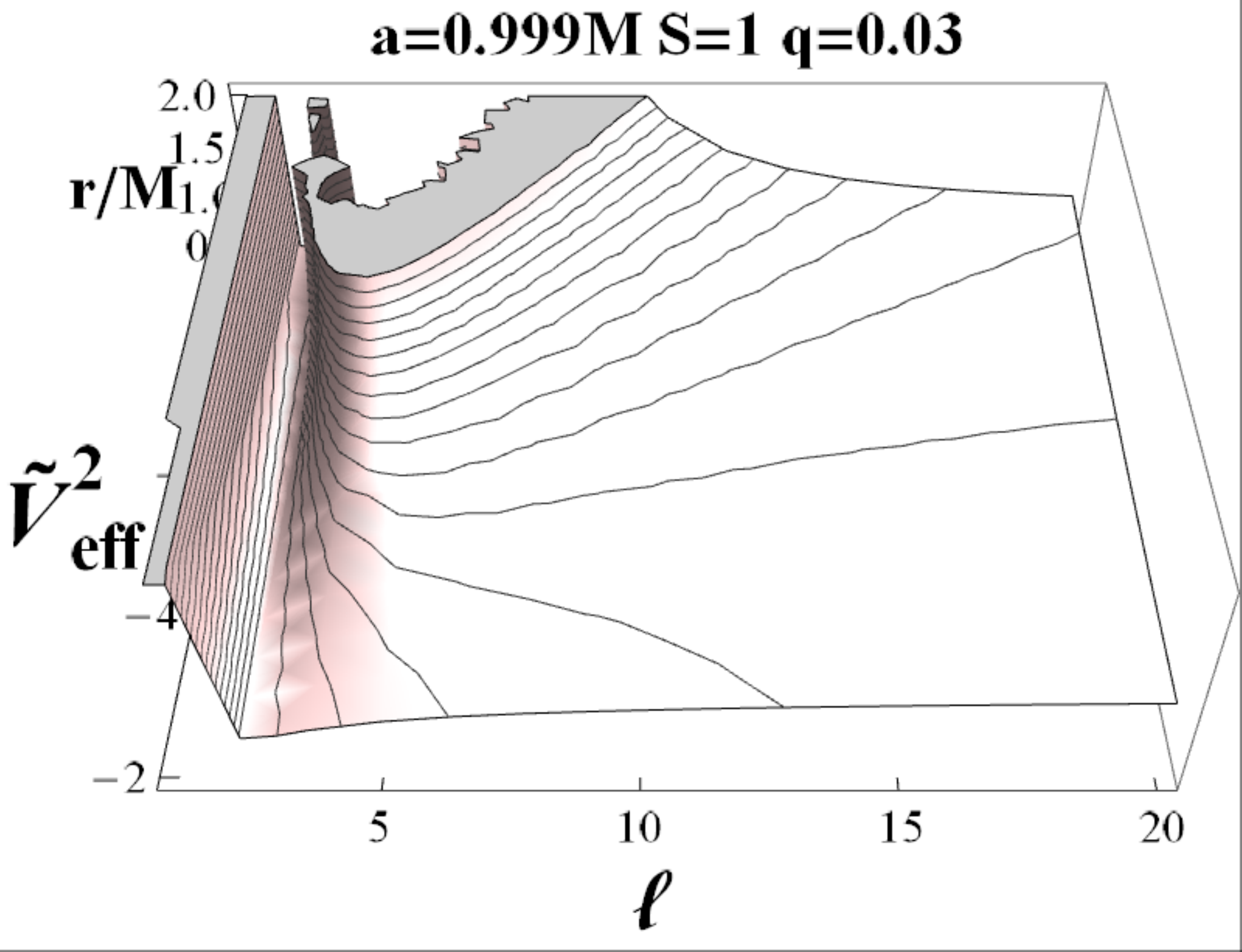}
\includegraphics[height=4cm, width=0.4\hsize,clip]{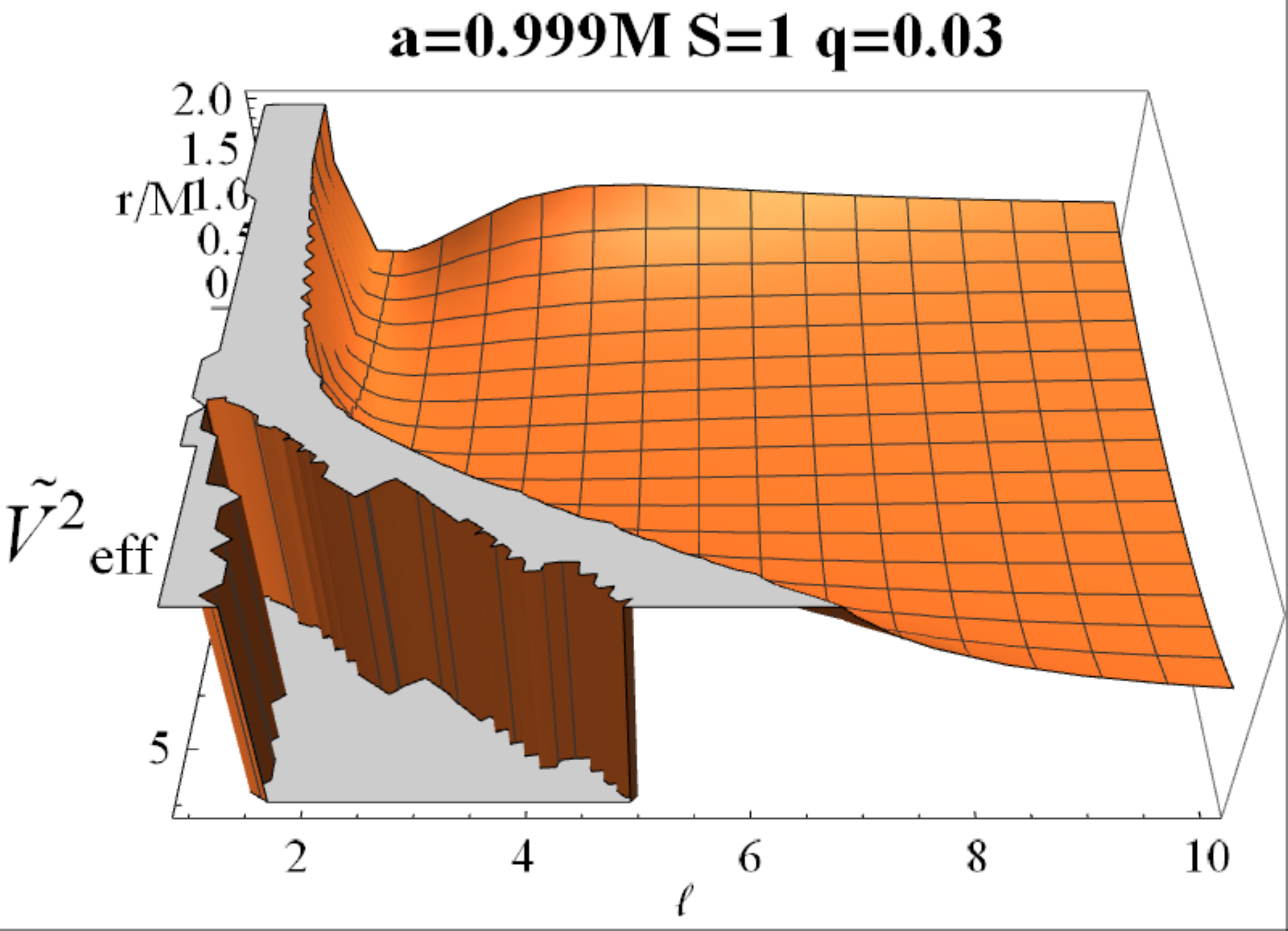}
\caption[font={footnotesize,it}]{Case $q\in]0,1[$ ($Q<0$):  Upper and middle panel-lines:  the closed Boyer surfaces at ${\rm{K}}^2$ and $\ell$ fixed  and associated  effective potential ($\Sa=1$). Disk center  $r_{\odot}$ and critical points $r_{\times}$ are also signed. Third line panel: density plot.  Double accretion configurations appear.
Bottom panels: different view of the effective potential as function of fluid specific angular momentum  $\ell$ and radius   $r/M$.  Double minima appear  for a restricted region  of parameter value $\ell<0$. $(x, y)$ are Cartesian coordinates. Black region is  $r<r_+$, $r_+$ is the Black Hole horizon.}
\label{Fig:Lyb-tody}
\end{figure*}
\section{{Notes on the   RAD instabilities}}\label{Sec:notes}
In this section we add some further  comments on the stability of the \textbf{RADs}  configurations  constituted by  tori  endowed with a toroidal magnetic field. We briefly   consider also the possible observational implications associated with the  instabilities.

\textbf{RAD} instabilities  should be treated in accordance with a global point of view where the  macrostructure  is considered  as a single, unique   disk  orbiting around  a central \textbf{SMBH}.
 In the   construction  of the \textbf{RAD} model, presented in \cite{ringed}, special attention has been given to the development of the \textbf{RAD} as  an whole, geometrically thin disk. In fact, the current interpretative framework  of the \textbf{BH}-accretion disk physics
 generally foresees the scenario of a \textbf{BH}-one disk system.  Consequently  we should consider the  possible situation of a ``\textbf{RAD} in disguise'', i.e.   a \textbf{RAD}  could be observed  as a geometrically  thin,  axis-symmetric disk, centered  on the equatorial plane on a  Kerr \textbf{SMBH},  with a ``knobby'' surface and     characterized by   a differential  rotation    with   peculiar optical properties. (Optical properties of a couple of orbiting tori are  expected to be investigated in a future work,  on the other hand X-ray emission are expected to shown the ringed  structure  in a discrete emission profile--see for example \cite{S11etal,KS10,Schee:2008fc,Schee:2013bya}.)
It is clear then that the instability of each \textbf{RAD} component  must reflect in an  inter-\textbf{RAD} disk activity.
More in general, the \textbf{RAD} instabilities have been   classified into  three main   processes:  \textbf{(i) } a destabilization of the system may arise after the  emergence an instability phase  of one component of the \textbf{RAD}, for example after an accretion phase of one torus onto the central  \textbf{BH} or  the proto-jet emission  which  is capable to destabilize the entire disk \citep{long,app}. This case however has been strongly constrained. As discussed in  Section\il\ref{Sec:MRADa}, in any  \textbf{RAD} the maximum number of accreting tori is $n_{\times}=2$,  occurring for the couple $\cc_{\times}^-<\cc_{\times}^+$, made by  an inner corotating   and outer counterrotating  torus  accreting  on the   \textbf{BH}. \textbf{(ii)} A \textbf{RAD} can be destabilized after  collision of a pair of  quiescent tori  of the agglomeration. Collision  may arise for example  after  growing of one torus \cite{dsystem,letter}. \textbf{(iii)}   In  the couple  $\pp^-<\cc_{\times}^+$, the  accretion phase of the  outer torus,    $\mathbf{(i)}$-instability,   and the collision emergence,  $\mathbf{(ii)}$-instability, can  combine  establishing a complex phase of \textbf{RAD} destabilization.
This situation has been discussed  in  \cite{ringed}, where \textbf{RAD} perturbative approaches  have also  been described. In  \cite{open},
the emergence of unstable tori have been detailed,  while further discussion on \textbf{RADs} as remnants of  \textbf{AGN} accretion periods are in  \cite{long}.
The particular case of the  emergence of collision for two \textbf{RAD} tori was considered in    \cite{dsystem}.
Interacting  tori and energetic of associated to these processes
     were investigated in  \cite{letter}. In this analysis the energy released during the  collision of two adjacent tori, $\cc_{\times}^-<\cc_{\times}^+$ or $\cc_{\times}^+<\cc^{\pm}$, has been evaluated.   The mass accretion rates,  the  luminosity at the cusps and other fundamental characteristics of the \textbf{BHs} accretion disk physic were also evaluated.
From the phenomenological viewpoint, the shift in paradigma  from the interpretative framework of the \textbf{BH}-disk interaction  to the \textbf{BH}-\textbf{RAD}  clearly opens a broad scenario of investigation focusing, on one side, on the special phenomena associated to the \textbf{RAD} instabilities , as the occurrence of double accretion and  its   after-dynamics, the inter disks proto-jet emission and  the screening tori; on the other side, the \textbf{RAD} model opens the possibility to  review  the main template of  analysis from a \textbf{SMBH}-disk  framework to a \textbf{SMBH}-\textbf{RAD} one.
More precisely, concerning  the  phenomenology connected to the (\textbf{i}), (\textbf{ii}) and (\textbf{iii}) instabilities,  the analysis in  \cite{letter}
suggests  that such phenomena can be associated with  release of high energy emissions.
Then from the point of view of the agglomerate, the collision instability can lead to different evolutive  paths  for the aggregate tori, depending on the initial conditions of the processes as  the torus rotation with respect to the black hole,  the range of variation of the mass of the torus  and of the  magnitude of the  specific angular momentum of the fluids.
A possibility consists in the formation of  in a single torus,  in fact canceling the \textbf{RAD} structure,  explaining mainly  in the first evolution phases of the  formation of the aggregate. We should also note that, as pointed out in \cite{dsystem}, an inner torus of the orbiting \textbf{RAD} couple may form as axially symmetric corotating toroidal disk after a first phase of formation of the outer aggregate component.

Conversely,  another possibility is the occurrence of a ``drying-feeding'' phase,  involving  interrupted stages of   accretion of one or two tori of a couple. In this case,
matter flows between the  two tori of the couple,  accretion being  interspersed with equilibrium phases, eventually giving raise to a series of  interrupted stages of accretion onto the central \textbf{SMBH}. This particular effect,  considered in \cite{dsystem,long} and detailed  in \cite{letter} can  represent a fitting environment for the  different phases of super-Eddington accretion  advocated as a mechanism to explain  the large masses observed in \textbf{SMBHs} at high redshift--see for example  \cite{apite1,apite2,apite3,Li:2012ts,Oka2017,Kawa,Allen:2006mh}.
In the case of a $\pp^-<\cc_{\times}^+<\cc^{\pm}$  system the inner, accreting or quiescent, torus
 can be an obscuring inner torus. Matter, from the outer counterrotating torus, impacts on the corotating inner one, which is screening the accretion  from the central \textbf{SMBH}. The possible evolutive paths  of such system  have been constrained  in the hydrodynamics case   using constraints on the variation ranges  of the \textbf{RAD} parameters and  on   Eq.\il(\ref{Eq:goood-da})--\cite{dsystem}.
 This treatment is semi-analytic, while the full evolution of the  collisional regime has to be considered apart.

%%%%%%%%%%%%%%%%%%%%%%%%%%%%%%%%%%%%%%%%%%%%%%%%%%%%%%%%%%%%%%%

More generally, each  torus  oscillation mode    will reflect on the  \textbf{RAD} structure  adding up to  those of other components of the  agglomerate, each torus will contribute  with  its own specific  characteristic. Eventually  this can be also related  to  \textbf{QPOs} emission--see \cite{Montero:2007tc}.
It is therefore necessary to consider the oscillations and instabilities associated with the each  component of the aggregate.  The introduction of a purely toroidal and   even  small magnetic field  (considering  the  magnetic pressure \emph{versus}  gas pressure as  defined by the $\beta$-parameter) can have influence on the development of these modes. This is relevant particular for the  the torus global  non-axis-symmetric modes,  because of   the generation of the  Magneto-Rotational Instability  (MRI) due to the  magnetic field and  fluid differential rotation.
Geometrically thick  discs are subjected to several oscillation modes: a first modes set  is  constituted by incompressible and  axis-symmetric  modes  which  correspond to  global  oscillations   for radial,  vertical  and   epicyclic  frequencies together with surface  gravity,
acoustic   and  internal    modes  which are  recovered from  the  so called  relativistic  Papaloizou-Pringle (\textbf{PP})
equation--see for example \cite{abrafra}.
On the other hand, the
Papaloizou-Pringle Instability
(\textbf{PPI}), is
a global, non-axis-symmetric instability which is able to transport angular momentum outwardly in the disk and therefore   able to finally trigger the  accretion.
The  global non-axis-symmetric hydrodynamic (\textbf{HD}) \textbf{PPI}
 implies also   the  formation of long-lasting,
large-scale structures that may be  also  tracer for such tori    in the
in the gravitational wave emission--see for example \cite{f-x-Kiuchi}.
We also note that  the presence of these  modes in complex structures such as those provided by the  \textbf{RAD} can be extremely intriguing, considering the possibility of the  emergence from distinct structures belonging to the aggregate, which are characterized  by    fluids with different  physical proprieties.

Accretion in \textbf{BH}  disks is   provided by an instability  process which is able to  trigger   the matter overflow in the torus.
In the  geometrically \textbf{HD} thick disks, the accretion process   is strictly interwoven  with the development of the \textbf{PP} instability:  the  mass  loss in  the   Roche  lobe  overflow  regulates  the  accretion  rate  in  the  innermost  part  of   torus. This   self-regulated process on one side
locally  stabilizes   the accreting torus from the   thermal  and  viscous   instabilities    and, on the other side, it   globally  stabilizes the torus from the \textbf{PPI}--\citep{Abramowicz-nATURE,Blaes}.
(Note also  that the amount of overflow    may  be also modulated by global disks oscillations.)
 In fact,  global instabilities  are affected by  the
  boundary
conditions assumed for the system. In the case of  \textbf{PPI} in  \textbf{RAD} accreting \textbf{HD} tori, for which  the disk inner and outer edges  are well defined and located,   the \textbf{PPI} is generally suppressed, stabilizing the disks by the accretion flow driven
by the  pressure forces across the cusp, $r_{\times}$, according to the mechanism considered in Section\il\ref{Sec:model}.
In the case of
  geometrically thick torus endowed with a (purely) toroidal  magnetic field,  considered  here with  the analytic  Komissarov solution, a series of recent analysis shows that  torus is violently prone to  develop the non-axisymmetric MRI in 3D which could  disturb this configuration on dynamical timescales--see \cite{Del-Zanna,Wielgus,Das:2017zkl} and \cite{Bugli}.
   The \textbf{PPI} hydrodynamic instability  is entangled with an  emerging  MRI which triggers  eventually predominant larger modes of oscillation (smaller length scales) with respect to typical \textbf{PPI} modes, and  creating a far richer and complex scenarios for  the torus equilibrium properties.
Therefore, the presence of a magnetic field contribution in the disk force balance leads  to a more complex situation where the \textbf{PPI} has to be considered  in a broader context. More generally, whether or  not the hydrodynamical  oscillation  modes   in MHD geometrically thick disks may  survive  such  global  instabilities  or  the  presence  of  a  weak  magnetic field  would strongly affects these, is still under  investigation.
The  linear development of the \textbf{PPI} can be  affected by the
presence of a magnetic field and by a combined growth of  the MRI.
 These two processes can coexist, enter into competition and  combine depending on local parameters of the model (strongness of the magnetic field as evaluated by $\beta$ parameter). Some studies  seem to suggest that under certain conditions on the strength of the magnetic field and other conditions on the torus onset, this situation can also be resolved in the \textbf{PPI} suppression by the MRI in  the relativistic
accretion disks.
 Using three-dimensional GRMHD simulations  it is also  studied the
interaction between the \textbf{PPI} and the MRI considering  an analytical magnetized equilibrium solution as initial condition \citep{Bugli}. In the \textbf{HD} tori, the
\textbf{PPI} selects the large-scale $m = 1$ azimuthal mode as the fastest growing and non-linearly dominant mode. In different works it is practically shown that  even a
weak toroidal magnetic field can lead to  MRI development which leads to   the suppression of the large-scale modes.
 Notice also that the   magneto-rotational instability  in the disks is  important because disks  can be   locally \textbf{HD} stable (according to  Rayleigh criterion), but they  are unstable for  \textbf{MHD} local  instability  which is
linear and
independent by the  field strength and orientation, and  growing up
   on dynamical time
scales. The torus  (flow) is \textbf{MHD} turbulent due to the MRI.
The MRI process
induces an  angular momentum transfer towards the outer region of  the  torus using  the
torque of the magnetic field lines.
In  the magnetized tori,  as the \textbf{RAD} tori considered here, the accretion is
triggered at much earlier times  then in the \textbf{HD} tori, and modes higher then the azimuthal  $m=1$ mode, typical of  \textbf{HD-PPI} tori, emerge together with $m=1$.  GRMHD investigations  show  generally an increase of   turbulent kinetic energy in the   earlier phases competing with the GRHD ones, consequently  accretion   is in fact  triggered by the Maxwell     stresses instead of the  \textbf{PPI}.
Furthermore, in the magnetized case there is a  broader range of excited frequencies with respect to the GRHD model.
Eventually  the fundamental
mechanism responsible for the onset of the \textbf{PPI} does not appear to be the predominant one or even to arise at all in the MHD torus.
In conclusion these works show that the  inclusion of a toroidal magnetic field could strongly  affect, even with a
sub-thermal  magnetic field, the  \textbf{PPI}.
Ultimately there are suggestions that the action of MRI suppresses the \textbf{PPI}  $m=1$ mode growth.
This may have a relevant  consequence in the double \textbf{RAD} system. MRI stabilizes the disks
to \textbf{PPI}  with  \textbf{MHD} turbulence.
Firstly in general MRI  is more effective and  faster in transport of  angular momentum across
the disk, and higher  accretion rates were proved to occur
in the magnetized models.
The evaluation of the accretion rates in  the GRHD double \textbf{RAD} systems has been carried out in \cite{letter}. The  emergence   of the MRI  suggests an
accentuation of the effects of the \textbf{(i)} and \textbf{ (iii)} instabilities,  whereas we do not expect the principal mechanisms to be changed but rather to accentuate those  phenomena connected with energy release and matter impact. Nevertheless these consideration  have to be dealt    with the constrains provided in  Section\il\ref{Sec:MRADa}.
Finally it should be noted that,  according to \cite{Fragile:2017lbx}, strong  toroidal  magnetic fields are rapidly suppressed  in this tori, in favor of weaker fields (decrease of $\beta$ parameter).
%i.e. where the angular velocity of the pattern matches the local orbital frequency
%of the disk.
%Other suggestion may even induce to suppression of PPI  (no waves
%beyond the corotation radius)
On the other hand, despite  these investigations   seem to converge towards  a quite clear picture of  the MRI-\textbf{PPI} interaction in geometrically thick disks, although indicative of what the situation could be in general,  more analysis  is definitely needed to  draw a more conclusive picture of this  interaction. The relative importance of  MRI and \textbf{PPI}  and  the interaction of two  processes depends in fact  on many factors and conditions.
In particular in the \textbf{RAD} scenario different factors can be determinant:
 the (turbolent) resistivity, the  emerging of a  dynamo effect, the study  for counterrotating  (retrograde) tori, the  disk self-gravity,  the gravitational interaction between the disk and the central Kerr \textbf{SMBH}  and   the runaway instability  are further aspects which may  contribute  importantly to the characterization of the ongoing  processes.

\section{Conclusions}\label{Sec:Discussion}
We studied   the effects of a toroidal magnetic field in the formation of multi-magnetized   accretion tori   in the  ringed accretion disks (\textbf{RADs}), orbiting around one central  supermassive Kerr Black Hole.
Results constitute  evidence of  a strict correlation between \textbf{SMBH} dimensionless spin,  fluid rotation  and magnetic fields in \textbf{RADs} formation and evolution towards instability. We showed  how the  central \textbf{BH} dimensionless spin, the presence of a  magnetic field and the relative fluid rotation  and the rotation  with respect the central attractor,  play a crucial role  in determining the accretion tori features.  Specifically, it is proved that toroidal  magnetic field and disks rotation are   strongly  related. This can  ultimately have a major influence in the \textbf{BH}-accretion disk systems, especially during the early stage of   tori  formation and the  final steps of evolutions towards the accretion onto the spinning  \textbf{BH},  a phase where  predominant   instabilities occur for  the accreting torus as well as for the \textbf{RAD} system. Noticeably,   we found that  only specific classes of   constrained tori, for     restrict  ranges of magnetic field parameters may  form    around   special \textbf{SMBHs}  belonging to  classes   identified by their dimensionless spin. This clearly   has huge implications for observational point of view, providing indications on the contexts  where to observe such configurations,  providing also    insight on the different stages of the  \textbf{BH} life interacting with its environment and the torus features. In section\il\ref{Sec:MRADa} we provided a detailed summary of the findings.
Only  for \textbf{BHs} with spin parameter $a\neq0$ and in a couple made by an outer counterrotating torus and inner corotating torus, a  double accretion occur, with the outer accreting matter impacting on the inner  ``screening'' disk,  which is  also accreting onto the central \textbf{BH}. This mechanism envisages  a special ``inter-disks'' activity with greater observational  potentiality and it  poses strict constraints of the current studies of \textbf{X}-ray emission  screening in \textbf{BH} environments, restricting strongly the situations where a screening effect from an orbiting inner tori can be considered\citep{Marchesi,Gilli:2006zi,Marchesi:2017did,Masini:2016yhl,DeGraf:2014hna,Storchi-Bergmann}. The possibility of tori collision  under the effect of the magnetic field  is also  enlighten  for  a system of  non-accreting couple and  for impact of matter inflow from the outer  onto the inner disk.
A modification of the tori rotation law (specific angular momentum), depending on the magnetic field is discussed. This has the advantage to provide a fairly small, though detailed, template of associated phenomenology, with special regard to situations where collisions and accretion occur.
The counterrotating  and $\ell$counterrotating cases  show  significantly that  the  toroidal magnetic field plays an essential role in determining  the disk structure and stability, showing that also a purely azimuthal field is capable to discriminate the  \textbf{RAD} features.
%In\cite{eff} for example it is shown that the toroidal magnetic component inside an accretion torus does
%not change the frequency of its oscillations significantly.
%The toroidal magnetic field plays a more important role in the early phases of the accretion process until the perturbed
%configuration finds a new equilibrium or disappears because of the runaway instability.
From a methodological point of view, the rewriting of Euler equations in the form of an equation with an  (general relativistic) effective potential allows us to  precisely estimate the  balance of each component of the forces regulating  the disk and  the \textbf{RADs} agglomerate.
In the choice of a particular set-up, especially for a   magnetized model, there is inevitably a level of arbitrariness in the specification of the model ending up to narrow the  range of situations where this can fit
to very specific contexts.
The single magnetized torus of \textbf{RAD} is  however widely used to fix up the initial configurations for  numerical integration of a broad
variety  of \textbf{GRMHD} models.
Parameterizing  the magnetic  field through the two parameters  $(q, \Sa) $,  we narrowed the range of parameter variation, relating the $\Sa$ parameter values to the system critical points.

In conclusion, the results of our analysis show that  the magnetic field has an important role in determining the {\textbf{RADs}} formation and instability. In this respect, as we already stressed in Section\il\ref{Sec:MRADa}, we should in fact  revisit the current  analysis of screened
X-ray emission,  by considering constraints provided here on the formation of an  inner screening torus.

DP acknowledges support from a Junior GACR Grant of the Czech Science Foundation No:16-03564Y.
This work has been developed in the framework of the CGW Collaboration
(www.cgwcollaboration.it).

\bsp	% typesetting comment
\label{lastpage}
\end{document}